\documentclass[structabstract]{aa}
\usepackage{graphicx}
\usepackage{txfonts}
\usepackage{natbib}
\bibpunct{(}{)}{;}{a}{}{,}
\usepackage{amssymb}
\usepackage{xcolor}
\usepackage{lipsum}
\usepackage{textcomp}
\usepackage[colorlinks=true,linkcolor=blue,citecolor=blue,
bookmarks=true,bookmarksopen=true,bookmarksnumbered=true,draft=false]{hyperref}
\usepackage{enumerate}

\newcommand{\xmm}{{\it XMM-Newton}}
\newcommand{\snr}{\object{MCSNR J0528$-$6727}}
\newcommand{\dem}{\object{DEM L205}}
\newcommand{\hour}{$^{\mathrm{h}}$}
\newcommand{\minute}{$^{\mathrm{m}}$}
\newcommand{\second}{$^{\mathrm{s}}$}

\begin{document}

   \title{Multi-frequency study of supernova remnants\\in the Large Magellanic
Cloud\,\thanks{Based on observations obtained with \xmm, an ESA science mission
with instruments and contributions directly funded by ESA Member States and
NASA.}
	}
   \subtitle{Confirmation of the supernova remnant status of DEM L205}

	\titlerunning{\dem, a new supernova remnant in the LMC}

   \author{P.  Maggi \inst{1}
	\and F.    Haberl \inst{1}
	\and L. M. Bozzetto \inst{2}
	\and M. D. Filipovi\'c \inst{2}
	\and S. D. Points \inst{3}
	\and Y.-H. Chu \inst{4}
	\and M.    Sasaki \inst{5}
	\and W.    Pietsch \inst{1}
 \and \\ R. A. Gruendl \inst{4}
	\and J.    Dickel \inst{6}
	\and R. C. Smith \inst{3}
	\and R.    Sturm \inst{1}
	\and E. J. Crawford \inst{2}
	\and A. Y. De Horta \inst{2}
	}

   \institute{Max-Planck-Institut f\"ur extraterrestrische Physik, Postfach
	1312, Giessenbachstr., 85741 Garching, Germany\\ \email{pmaggi@mpe.mpg.de}
	\and
	University of Western Sydney, Locked Bag 1797, Penrith South DC, NSW 1797,
	Australia
	\and
	Cerro Tololo Inter-American Observatory, National Optical
	Astronomy Observatory, Cassilla 603 La Serena, Chile 
	\and
	Astronomy Department, University of Illinois, 1002 West Green Street,
	Urbana, IL 61801, USA
	\and
	Institut f\"ur Astronomie und Astrophysik T\"ubingen, Universit\"at
	T\"ubingen, Sand 1, 72076 T\"ubingen, Germany
	\and
	Physics and Astronomy Department, University of New Mexico, MSC 07-4220,
	Albuquerque, NM 87131, USA
    }

   \date{Received 30 May, 2012; accepted 23 July, 2012}

  \abstract
  {The Large Magellanic Cloud (LMC) is an ideal target for the study of
an unbiased and complete sample of supernova remnants (SNRs). We started an
X-ray survey of the LMC with \xmm, which, in combination with observations at
other wavelengths, will allow us to discover and study remnants that are either
even fainter or more evolved (or both) than previously known.}
   {We present new X-ray and radio data of the LMC SNR candidate \dem, obtained
by \xmm\ and ATCA, along with archival optical and infrared observations.}
   {We use data at various wavelengths to study this object and its complex
neighbourhood, in particular in the context of the star formation activity, past
and present, around the source. We analyse the X-ray spectrum to derive some
remnant's properties, such as age and explosion energy.}
  {Supernova remnant features are detected at all observed wavelengths\,: soft
and extended X-ray emission is observed, arising from a thermal plasma with a
temperature $kT$ between 0.2 keV and 0.3 keV. Optical line emission is
characterised by an enhanced [\ion{S}{ii}]-to-H$\alpha$ ratio and a shell-like
morphology, correlating with the X-ray emission. The source is not or only
tentatively detected at near-infrared wavelengths (shorter than 10 $\mu$m), but
there is a detection of arc-like emission at mid and far-infrared wavelengths
(24 and 70 $\mu$m) that can be unambiguously associated with the remnant. We
suggest that thermal emission from dust heated by stellar radiation and shock
waves is the main contributor to the infrared emission. Finally, an extended and
faint non-thermal radio emission correlates with the remnant at other
wavelengths and we find a radio spectral index between $-$0.7 and $-$0.9, within
the range for SNRs. The size of the remnant is $\sim 79 \times 64$ pc and we
estimate a dynamical age of about 35\,000 years.}
   {We definitely confirm \dem\ as a new SNR. This object ranks amongst the
largest remnants known in the LMC. The numerous massive stars and the recent
outburst in star formation around the source strongly suggest that a
core-collapse supernova is the progenitor of this remnant.}


   \maketitle

\begin{table*}[t]
\caption{Details of the \xmm\ observations}
\label{table_info}
\centering
\begin{tabular}{c c c c c c c c c}
\hline\hline
\noalign{\smallskip}
ObsId & Obs. start date & \multicolumn{2}{c}{Central coordinates (J2000)}
& Filter \tablefootmark{a} & \multicolumn{3}{c}{Total\,/\,filtered exposure
time
(ks) \tablefootmark{b}} & Off-axis \\
 &  & RA & DEC &  pn\,/\,MOS1\,/\,MOS2 & pn & MOS1 & MOS2 & angle
\tablefootmark{c}\\
\noalign{\smallskip}
\hline
\noalign{\smallskip}
0671010101 & 2011 Dec 19 & 05\hour\,29\minute\,55.7\second &
$-$67\degr\,26\arcmin\,14\arcsec & T\,/\,M\,/\,M & 25.0\,/\,20.1 &
26.6\,/\,21.7 & 26.6\,/\,21.7 & 8.8 \\
0071940101 & 2001 Oct 31 & 05\hour\,26\minute\,04.9\second &
$-$67\degr\,27\arcmin\,21\arcsec & T\,/\,T\,/\,T & 27.6\,/\,26.8 &
31.9\,/\,31.2 & 31.9\,/\,31.2 & 13.7 \\
\noalign{\smallskip}
\hline
\end{tabular}
\tablefoot{
\tablefoottext{a}{T\,: Thin ; M\,: Medium.}
\tablefoottext{b}{Performed duration (total) and useful (filtered) exposure
times, after removal of high background intervals.}
\tablefoottext{c}{Angle in arcmin between the centre of the pn detector and the
centre of the X-ray source (as defined in Sect.\,\ref{data_xray_image}).}
}
\end{table*}

%
\section{Introduction}
\label{introduction}

Supernova remnants (SNRs) are an important class of objects, as they contribute 
to the energy balance and chemical enrichment and mixing of the interstellar
medium (ISM). However, in our own Galaxy, distance uncertainties and high
absorption inhibit the construction of a complete and unbiased sample of SNRs.
On the other hand, the Large Magellanic Cloud (LMC) offers a target with a low
foreground absorption at a relatively small distance of $\sim$ 50 kpc
\citep{2008MNRAS.390.1762D}.

Furthermore, the broad multi-frequency coverage of the LMC, from radio to
X-rays, allows for easier detection and classification of SNRs, which is
most usually done using three signatures\,: thermal X-ray emission in the
(0.2--2~keV) band, optical line emission with enhanced [\ion{S}{ii}] to
H$\alpha$ ratio \citep[$\gtrsim$ 0.4,][]{1973ApJ...180..725M}, and non-thermal
(synchrotron) radio-continuum emission, with a typical spectral index of $\alpha
\sim - 0.5$ (using $S \propto \nu\,^{\alpha}$, where $S$ is the flux density and
$\nu$ the frequency), although $\alpha$ can have a wide range of values
\citep{1998A&AS..130..421F}. Nevertheless, the interstellar environment in which
the supernova (SN) exploded strongly affects the subsequent evolution of the
remnant, so that some SNRs do not exhibit all three conventional signatures
simultaneously \citep{1997AJ....113.1815C}.

\begin{figure}[!b]
	\centering
	\includegraphics[width=\hsize]{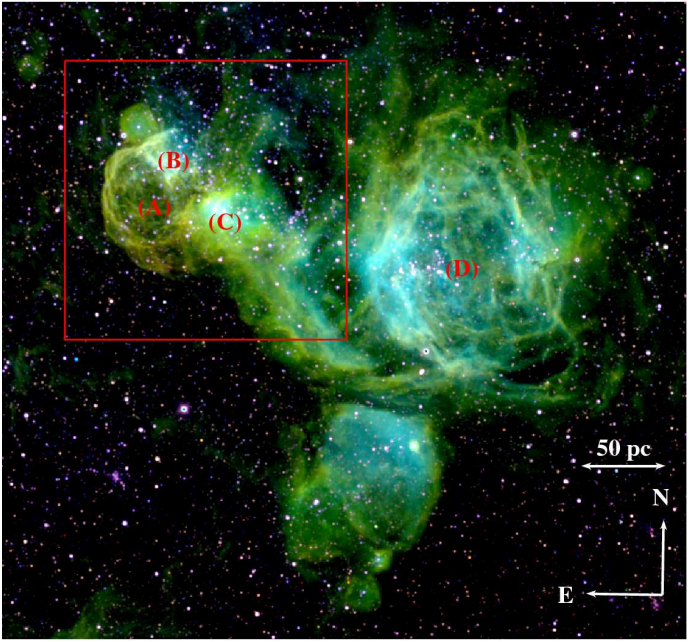}
	\caption{The giant \ion{H}{ii} complex \mbox{\object{LHA 120-N 51}} in the
light of  [\ion{S}{ii}] (red), H$\alpha$ (green), and [\ion{O}{iii}] (blue), all
data from MCELS (see Sect.\,\ref{observations_optical}). The red box delineates
the area shown in Fig.\,\ref{fig_rgb_image}. Noticeable substructures are\,: DEM
L205 (A), the SNR candidate analysed in this paper; N51A (B) and N51C (C, also
named DEM L201), two \ion{H}{ii} regions also seen in the radio and the IR; the
SB N51D, or DEM L192, in (D).}
	\label{fig_rgb_N51}
\end{figure}

In this paper, we present new X-ray and radio-continuum observations (with \xmm\
and ATCA) of the LMC SNR candidate \dem. Archival optical and infrared (IR)
observations are analysed as well. The source lies in a very complex
environment, at the eastern side of the \ion{H}{ii} complex \mbox{\object{LHA
120-N 51}} \citep[in the nebular notation of][] {1956ApJS....2..315H}. It was
classified as a ``possible SNR'' by \citet*{1976MmRAS..81...89D} (from which the
identifier ``DEM'' is taken) based on its optical shell-like morphology. In
X-rays, the catalogue of \emph{ROSAT}'s PSPC sources in the LMC
\citep{1999A&AS..139..277H} includes \object{[HP99] 534}, located within the
extent of \object{DEM L205}. However, the short exposure and large off-axis
position prevented any classification of the X-ray source.
\citet{2001ApJS..136..119D} classified \object{DEM L205} as a superbubble (SB)
with an excess of X-ray emission.

In our new and archival observations, we detected the object at all wavelengths.
The subsequent analyses allowed us to confirm the SNR nature of the source and
estimate some of its parameters. The paper is organised as follows: in
Sect.\,\ref{observations}, we present our new X-ray and radio-continuum
observations, and archival optical and IR data. The X-ray, radio, and IR data
analyses are detailed in Sect.\,\ref{data}. We discuss the implications of our
multi-frequency study in Sect\,\ref{discussion}, and we give our conclusions in
Sect.\,\ref{conclusions}.

\begin{figure*}[t]
	\begin{center}
	\includegraphics[width=0.49\hsize]{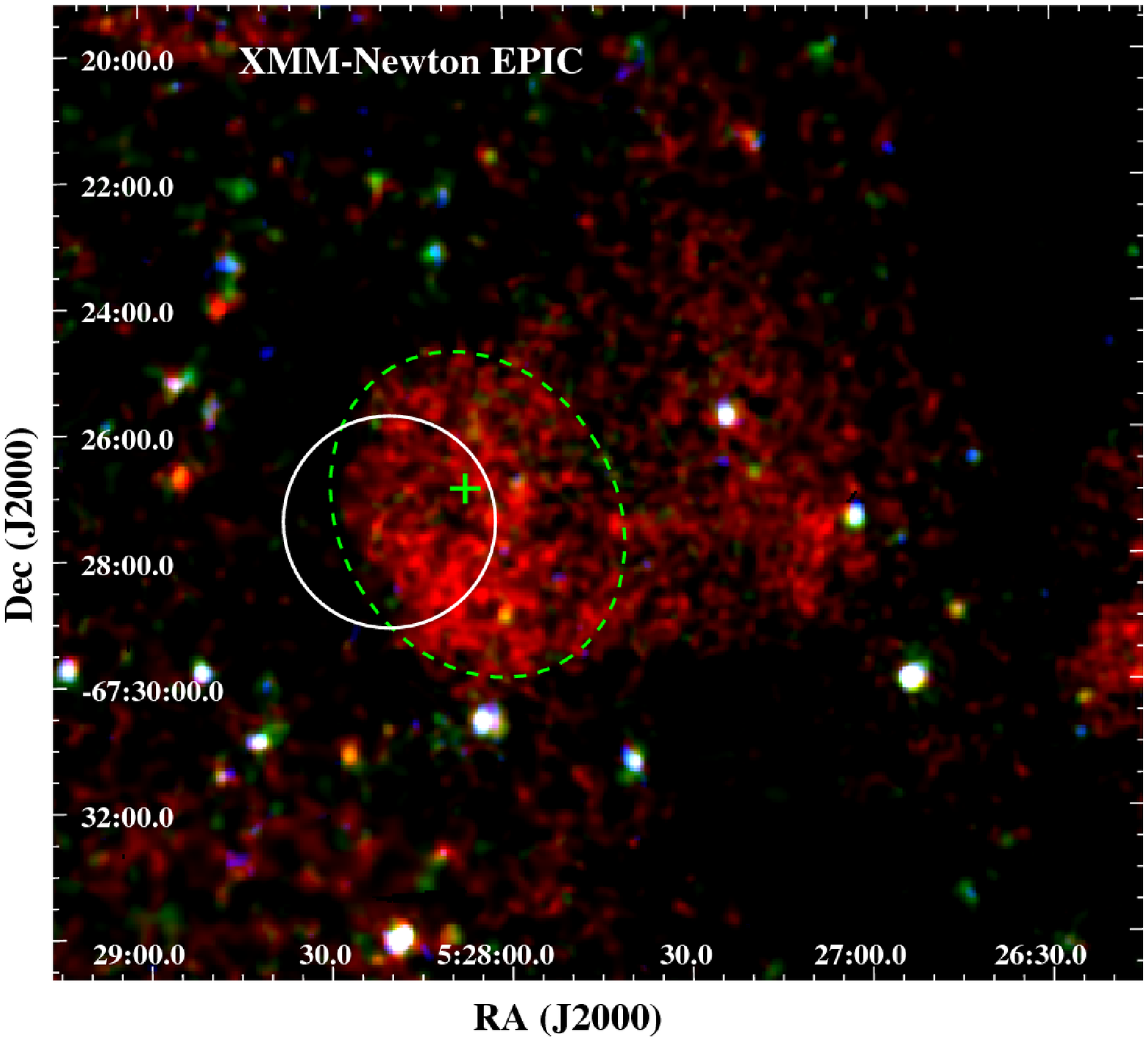}
	\includegraphics[width=0.49\hsize]{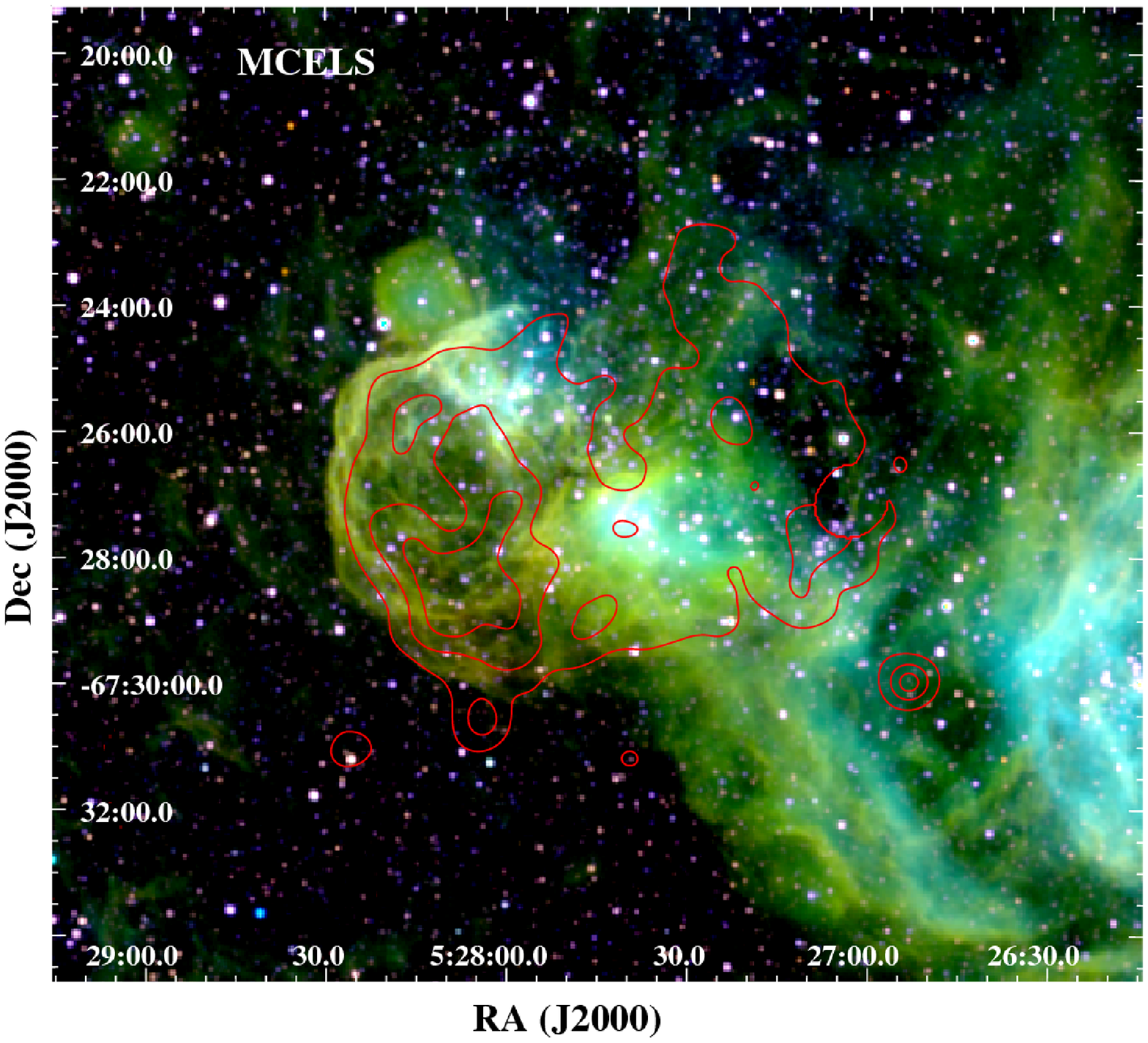}

	\includegraphics[width=0.49\hsize]{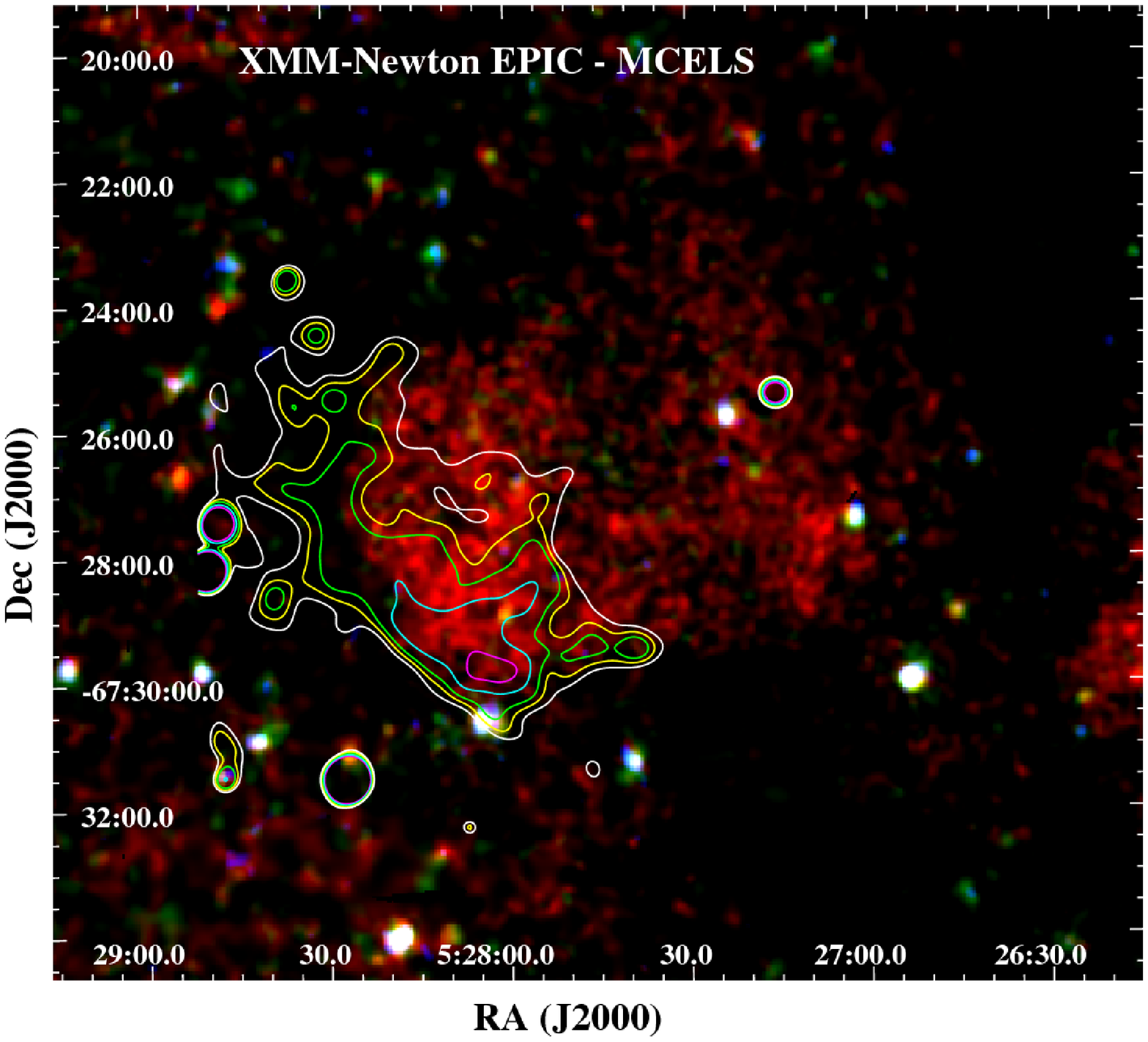}
	\includegraphics[width=0.49\hsize]{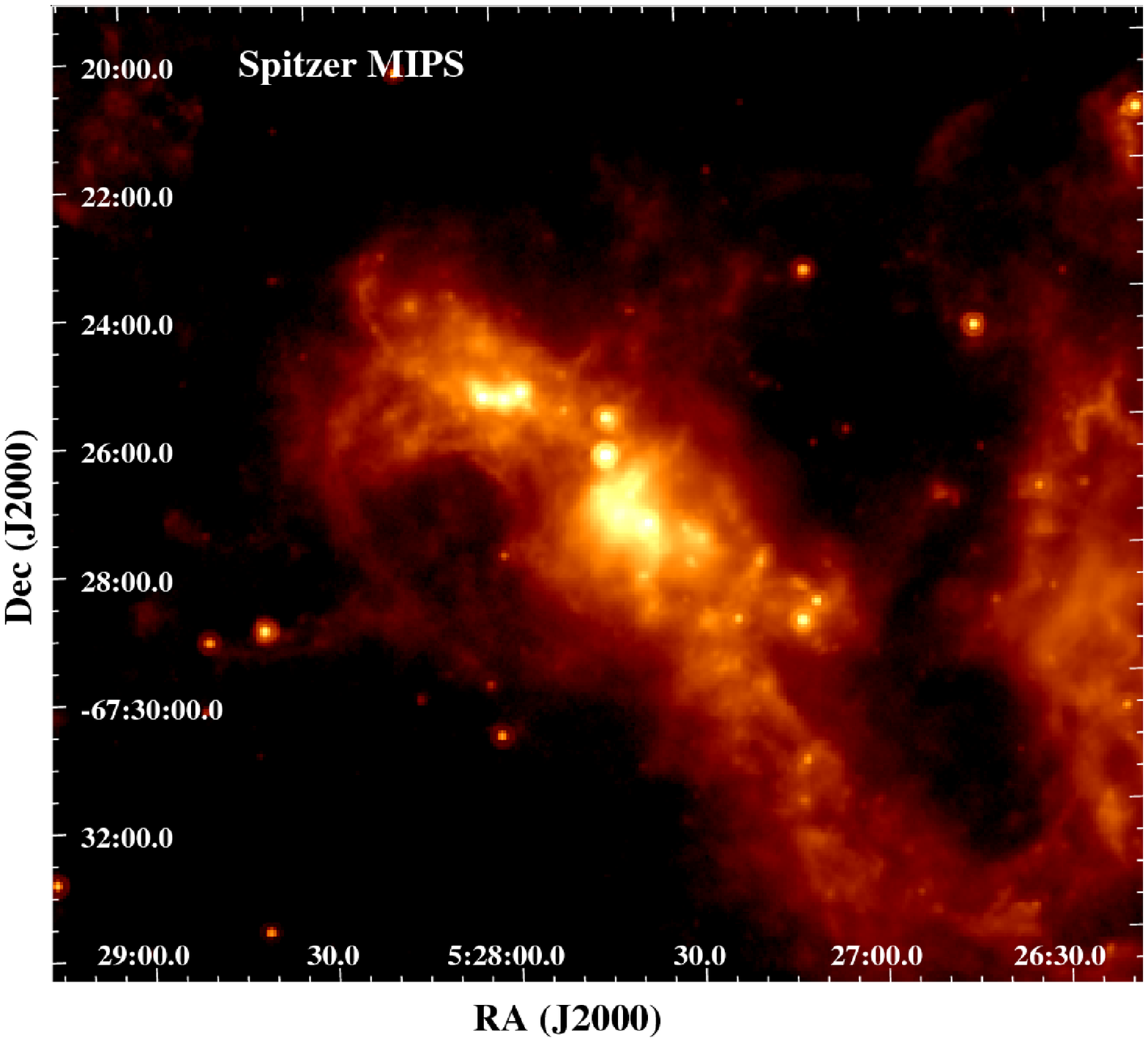}
	\end{center}
	\sidecaption
\caption{A multicolour view of \dem.
\emph{Top left}\,: X-ray colour image of the remnant, combining all EPIC
cameras. Data from two overlapping observations are combined and smoothed (see
Sect.\,\ref{data_xray_image} for details). The red, green, and blue components
are soft, medium, and hard X-rays, as defined in the text. The white circle is
the 90\,\% confidence error of the \object{[HP99] 534} position and the green
cross is the central position of \object{DEM L205}. The green dashed ellipse
(defined in Sect.\,\ref{data_xray_image}) encompasses  the X-ray emission and is
used to define the nominal centre and extent of the remnant.
\emph{Top right}\,: The same region of the sky in the light of [\ion{S}{ii}]
(red), H$\alpha$ (green), and [\ion{O}{iii}] (blue),where all data are from the
MCELS. The soft X-ray contours from the top left image are overlaid.
\emph{Bottom left}\,: Same EPIC image as above but with
[\ion{S}{ii}]-to-H$\alpha$ ratio contours from MCELS data. Levels are (inwards)
0.4, 0.45, 0.5, 0.6, and 0.7.
\emph{Bottom right}\,: The remnant as seen at 24 $\mu$m by \emph{Spitzer} MIPS.
Optical and IR images are displayed logarithmically.}
		\label{fig_rgb_image}
\end{figure*}

%
\section{Observations and data reduction}
\label{observations}

\subsection{X-rays}
\label{observations_xray}

\dem\ was in the field of view of the European Photon Imaging Camera
(EPIC) X-ray instrument, comprising a pn CCD imaging camera
\citep{2001A&A...365L..18S} and two MOS CCD imaging cameras
\citep{2001A&A...365L..27T}, during the first pointing of our recently started
LMC survey with \mbox{\xmm}. The 28~ks observation (ObsId 0671010101) was
carried out on 19 December 2011. The EPIC cameras operating in full-frame mode
were used as the prime instruments. We used the XMM SAS \footnote{Science
Analysis Software, \url{http://xmm.esac.esa.int/sas/}} version 11.0.1 for the
data reduction. After the screening of high background-activity intervals, the
useful exposure times for pn and MOS detectors were $\sim$ 20 and 22~ks,
respectively. 

An archival \xmm\ observation (ObsId 0071940101, pointing at the LMC SB N\,51D)
includes the remnant in the field of view, at a larger off-axis angle.
None of the EPIC cameras covered the remnant to its full extent. We used data
from this observation only for the purpose of imaging but not spectrometry. This
gives us a longer exposure time, particularly in the western part of the
remnant, which is covered by all cameras in both observations. In Table
\ref{table_info} we list the details of the observations.

Images and exposure maps were extracted in the standard \xmm\ energy bands
\citep[see Table~3 in][]{2009A&A...493..339W} for all three cameras. Single
and double-pixel events (\texttt{PATTERN} = 0 to 4) from the pn detector were
used. In the softest band (0.2 -- 0.5 keV), only single-pixel events were
selected to avoid the higher detector noise contribution from the double-pixel
events. All single to quadruple-pixel (\texttt{PATTERN} = 0 to 12) events from
the MOS detectors were used. We performed a simultaneous source detection on
images in all five energy bands of all three instruments, using the SAS task
\emph{edetectchain}.

We then subtracted the detector background taken from filter wheel closed data,
scaled by a factor estimated from the count rates in the corner of the images
not exposed to the sky. MOS and pn data were merged and we applied a mask to
remove bad pixels. Images from the two observations were merged and adaptatively
smoothed, using Gaussian kernels with a minimum full width at half maximum
(FWHM) of 10\arcsec. Kernel sizes were computed at each position in order to
reach a typical signal-to-noise ratio of five. Finally, we divided the smoothed
images by the vignetted exposure maps.

\subsection{Radio}
\label{observations_radio}

We observed \dem\ with the Australia Telescope Compact Array (ATCA) on the
15 and 16 November 2011 at wavelengths of 3~cm and 6~cm (9000~MHz and 5500~MHz),
using the array configuration EW367. We excluded baselines formed with the sixth
antenna, leaving the remaining five antennae to be arranged in a compact
configuration. More information about this observation and the data reduction
can be found in \citet{2012A&A...540A..25D}. Our 6~cm
observations were merged with those from
\citet{2005AJ....129..790D,2010AJ....140.1567D}. In addition, we made use of the
36~cm Molonglo Synthesis Telescope (MOST) unpublished mosaic image as described
by \citet{1984AuJPh..37..321M} and an unpublished 20~cm mosaic image from
\citet{2007MNRAS.382..543H}. Beam sizes were
46.4\arcsec\,$\times$\,43.0\arcsec\ for the 36 and 20~cm images. Our 6~cm image
beam size was 41.8\arcsec\,$ \times$\,28.5\arcsec, with a position angle of
49.6\degr.

\begin{table*}[t]
\caption{X-ray spectral results for \dem}
\label{table_spectral_results}
\centering
\begin{tabular}{l c c c c c c c}
\hline
\hline
\noalign{\smallskip}
\multicolumn{8}{c}{Background model best-fit parameters}\\
\noalign{\smallskip}
\hline
\noalign{\smallskip}
Model & N$_{H\mathrm{\ Gal}}$\tablefootmark{a} &
$kT_{\mathrm{Halo}}$ & EM$_{\mathrm{Halo}}$\tablefootmark{b} &
$\Gamma _{\mathrm{XRB}}$\tablefootmark{a,\,}\tablefootmark{c}&
A$_{\mathrm{XRB}}$& 
$\Gamma _{\mathrm{SPC}}$ & A$_{\mathrm{SPC}}$ \\
 & ($10^{20}$ cm$^{-2}$) & (eV) & ($10^{57}$ cm$^{-3}$) & & & & \\
\noalign{\smallskip}
\hline
\noalign{\smallskip}
vapec   & 5.9 & 201$_{-9} ^{+20}$ & 5.2$_{-0.8} ^{+0.8}$ &
1.46 & 2.68$_{-1.3} ^{+1.4} \times 10^{-5}$ & 
0.78$_{-0.06} ^{+0.05}$ & 8.48$_{-0.9} ^{+0.9} \times 10^{-2}$ \\
vpshock & 5.9 & 203$_{-12} ^{+15}$ & 5.1$_{-0.8} ^{+0.7}$ &
1.46 & 2.39$_{-1.5} ^{+1.1} \times 10^{-5}$ &
0.79$_{-0.06} ^{+0.05}$ & 8.69$_{-1.0} ^{+1.0} \times 10^{-2}$ \\
vsedov  & 5.9 & 205$_{-12} ^{+13} $& 5.0$_{-0.5} ^{+0.6}$ &
1.46 & 2.02$_{-0.7} ^{+1.4} \times 10^{-5}$ & 
0.79$_{-0.02} ^{+0.03}$ & 8.93$_{-.04} ^{+0.4} \times 10^{-2}$ \\
\noalign{\smallskip}
\noalign{\smallskip}
\hline
\hline
\noalign{\smallskip}
\noalign{\smallskip}
\multicolumn{8}{c}{Source models best-fit parameters}\\
\noalign{\smallskip}
\hline
\noalign{\smallskip}
Model & N$_{H\mathrm{\ LMC}}$\tablefootmark{a} & $kT$ & $\tau$ & EM & 
12 + log(O/H) & 12 + log(Fe/H) & $\chi ^2 $ / dof\\
&($10^{20}$ cm$^{-2}$) &(eV)&($10^{12}$ s\,cm$^{-3}$) &($10^{57}$ cm$^{-3}$)& \\
\noalign{\smallskip}
\hline
\noalign{\smallskip}
vapec   & 0 & 251$_{-18} ^{+18}$ & --- & 20.6$_{-3.0} ^{+3.0}$ & 
8.22$_{-0.11} ^{+0.10}$ & 6.57$_{-0.33} ^{+0.25}$ & 452.96 / 490 \\
vpshock & 0 & 257$_{-33} ^{+60}$ & 4.21 ($>1.02$) & 22.0 & 
8.10$_{-0.11} ^{+0.11}$ & 6.86$_{-0.31} ^{+0.27}$ & 446.09 / 489 \\
vsedov  & 0 & 203$_{-20} ^{+72}$ & 50.0 ($>1.90$) & 24.5 & 
8.16$_{-0.10} ^{+0.12}$ & 6.84$_{-0.30} ^{+0.25}$ & 447.44 / 489 \\
\noalign{\smallskip}
\hline
\end{tabular}
\tablefoot{The top panel lists the best-fit parameters of the background model
and the bottom panel shows the parameters of the source models (details are in
Sect.\,\ref{data_xray_spectrum}). Errors are given at the 90\,\% confidence
level. We required the emission measures of the vpshock and vsedov models to be
in the 99\,\% CL range of EM obtained with the vapec model (see
Sect\,\ref{data_xray_spectral_results}), and this gave us the range of errors
for $kT$ and $\tau$ of the vpshock and vsedov models . The $\chi^2$ and
associated degrees of freedom (dof) are also listed.\\
\tablefoottext{a}{Fixed parameter (see text for details).}
\tablefoottext{b}{Emission measure $\int n_e n_H dV$.}
\tablefoottext{c}{$\Gamma _i$ and A$_i$ are the spectral indices and
normalisations of the power-law component $i$, where $i$ is either the X-ray
background (XRB) or soft proton contamination (SPC). A$_i$ are given in
photons\,keV$^{-1}$\,cm$^{-2}$\,s$^{-1}$ at 1 keV.}
}
\end{table*}

\subsection{Optical}
\label{observations_optical}
We used data from
the Magellanic Clouds Emission Line Survey \citep[MCELS, \emph{e.g.}] []
{2000ASPC..221...83S}. A 8\degr\,$\times$\,8\degr\ region centred on the LMC
was observed with the 0.6~m Curtis Schmidt telescope from the University of
Michigan/Cerro Tololo Inter-American Observatory (CTIO), using the three
narrow-band filters [\ion{S}{ii}]$\lambda\lambda$6716,\,6731 \AA, H$\alpha$, and
[\ion{O}{iii}]$\lambda$5007 \AA\ and matching red and green continuum filters.
We flux-calibrated and combined all MCELS data covering the SNR, with a pixel
size of 2\arcsec\,$\times$\,2\arcsec. The [\ion{S}{ii}] and H$\alpha$ images
were stellar continuum-subtracted to produce a map of [\ion{S}{ii}]/H$\alpha$,
which is an efficient criterion for distinguishing SNRs from \ion{H}{ii}
regions, where the ratio is typically $>$ 0.4 and $\lesssim$ 0.1, respectively
\citep{1985ApJ...292...29F}. The X-ray composite image and
[\ion{S}{ii}]/H$\alpha$ contours from these data are shown in
Fig.\,\ref{fig_rgb_image}. In addition, we present an unpublished
higher-resolution H$\alpha$ image in Fig.\,\ref{fig_environment} (pixel size of
1\arcsec\,$\times$\,1\arcsec), which was obtained as part of the ongoing MCELS2
program with the MOSAIC\,{\sc ii} camera on the Blanco \mbox{4-m} telescope at
the CTIO.

\subsection{Infrared}
\label{observations_infrared}
During the SAGE survey \citep{2006AJ....132.2268M}, the \emph{Spitzer Space
Telescope} observed a 7\degr\,$\times$\,7\degr\ area in the LMC with the
Infrared Array Camera \citep[IRAC;][] {2004ApJS..154...10F} in its 3.6, 4.5,
5.8, and 8 $\mu$m bands, and with the Multiband Imaging Photometer
\citep[MIPS;][]{2004ApJS..154...25R} in its 24, 70, and 160~$\mu$m bands. To
study the
IR emission of the source and its environment, we retrieved the IRAC and MIPS
mosaiced, flux-calibrated (in units of MJy\,sr$^{-1}$) images processed by the
SAGE team. The pixel sizes are 0.6\arcsec\ for all IRAC wavelengths and
2.49\arcsec\ and 4.8\arcsec\ for 24 $\mu$m and 70 $\mu$m MIPS data,
respectively.

%
\section{Data analysis and results}
\label{data}

\subsection{X-rays}
\label{data_xray}

\subsubsection{Imaging}
\label{data_xray_image}
We created composite images, using the energy ranges 0.2--1~keV for the red
component, 1--2~keV for the green, 2--4.5~keV for the blue. The X-ray image is
shown in Fig.\,\ref{fig_rgb_image}. In addition to soft diffuse emission and
many point sources, an extended soft source is clearly seen. This source
correlates with the positions of \object{[HP99] 534} and \object{DEM L205}. The
images alone already show that the source has hardly any emission above 1 keV.

The X-ray emission can be clearly delineated by an ellipse centred at RA =
05\hour\,28\minute\,05\second\ and DEC = $-$67\degr\,27\arcmin\,20\arcsec, with
a position angle of 30\degr\ (with respect to the north and towards the east,
see Fig.\,\ref{fig_rgb_image}). The major and minor axes have sizes of
5.4\arcmin\ and 4.4\arcmin, respectively. At a distance of 50 kpc, this
corresponds to an extent of $\sim$ 79 $\times$ 64 pc. We note that the eastern
and southern boundaries of the X-ray emission are more clearly defined than the
western and northern ones. We discuss this issue in
Sect.\,\ref{discussion_multi_frequency}.

\begin{figure*}[t]
	\centering
	\includegraphics
[bb= 69 0 580 700,clip,angle=-90,width=0.490\hsize]{vsedov_label.ps}
	\includegraphics
[bb= 50 34 574 725,clip,angle=-90,width=0.505\hsize]{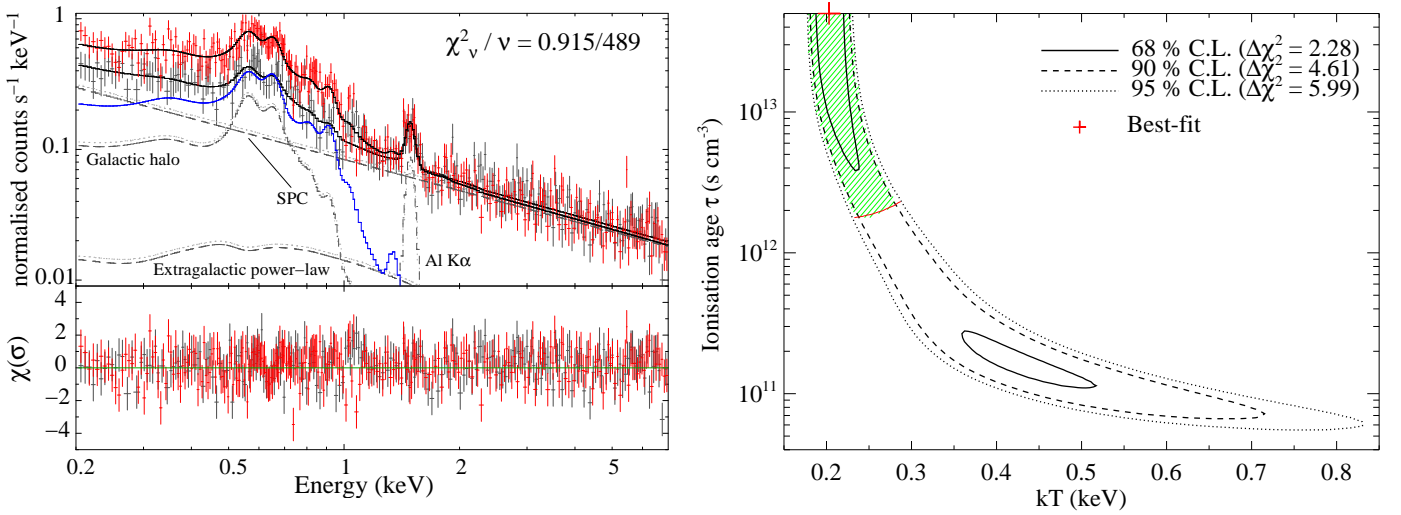}
\caption{\emph{Left}\,: EPIC-pn spectrum of \dem. The spectra in the
background and source regions (grey and red data points, respectively) are
modelled simultaneously. The background model components are shown by the dashed
lines and labelled. The Sedov model used for the remnant is shown by the blue
solid line. Residuals are shown in the lower panel in terms of $\sigma$.\newline
\emph{Right}\,: The $kT$ -- $\tau$ parameter plane for the Sedov model. The 68,
90, and 95\,\% CL contours are shown by the solid, dashed, and dotted black
lines, respectively. The formal best-fit, occurring at the upper limit of the
ionisation ages of the XSPEC model ($5\times 10^{13}$ s\,cm$^{-3}$) is marked by
the red plus sign. The red line shows the 99\,\% CL lower contour of emission
measure obtained with the APEC model. The green hatching indicates the region
where $\Delta \chi ^2 < 4.61$ and EM is in the 99\,\% CL range of the APEC model
(see Sect.\,\ref{data_xray_spectral_results}).
}
\label{fig_xray_spectrum}
\end{figure*}

\subsubsection{Spectral fitting}
\label{data_xray_spectrum}

We created a vignetting-weighted event list to take into account the effective
area variation across the source extent. The spectrum was extracted from a
circular region with a radius of 3\arcmin\ and the same centre as the ellipse
defined in Sect.\,\ref{data_xray_image}. A nearby region of the same size, free
of diffuse emission, was used to extract a background spectrum. Point sources
were excluded from the extraction regions. Spectra were rebinned with a minimum
of 30 counts per bin to allow the use of the $\chi ^2$-statistic. The
spectral-fitting package XSPEC \citep{1996ASPC..101...17A} version 12.7.0u was
used to perform the spectral analysis.

Because of the low surface brightness of the source, the spectrum of the source
region (background + source) contains a relatively low number of counts.
Constraining the parameters of the source model was therefore challenging. The
spectra were extracted from different regions of the detector (even from
different CCD chips) that have different responses. Simply subtracting the
background spectrum would have led to a further loss in the statistical
quality of the source spectrum, hindering the spectral fitting of the source
emission. A way to prevent this was to estimate the background using a
physically motivated model (although any model fitting the data could in
principle be used), and then fit the spectra of the source and background
regions simultaneously.

For the X-ray background, we used a three-component model, as in
\citet{2010ApJS..188...46K}, which consists of \emph{i)} an unabsorbed thermal
component for the Local Hot Bubble (LHB), \emph{ii)} an absorbed thermal
component to model the Galactic halo emission, and \emph{iii)} an absorbed
power-law to account for non-thermal, unresolved extragalactic background. The
spectral index was fixed to 1.46 \citep{1997MNRAS.285..449C}. We used
photoelectric absorption with the cross-sections taken from
\citet{1992ApJ...400..699B} and assumed the elemental abundances of
\citet{2000ApJ...542..914W}. 

In addition to this X-ray background model, an instrumental fluorescent line of
Al K$\alpha$, with $E = 1.49$ keV, and a soft proton contamination (SPC) term
were also included. The SPC was modelled by a power-law \emph{not} convolved
with the instrumental response, which is appropriate for photons but not for
protons \citep{2008A&A...478..575K}.

Three models were used for the emission of the remnant\,: a thermal plasma,
using the Astrophysical Plasma Emission Code (APEC), a plane-parallel shock, and
a Sedov model \citep{2001ApJ...548..820B}, called vapec, vpshock, and vsedov in
XSPEC (where the prefix ``v'' indicates that abundances can vary). The vsedov
model computes the X-ray spectrum of an SNR in the Sedov-Taylor stage of its
evolution. The parameters of the Sedov model are the mean shock temperature
$T_s$, the postshock electron temperature $T_{es}$ and the ionisation timescale
$\tau_0$, defined as the product of the electron density behind the shock front
and the remnant's age. As \citeauthor{2001ApJ...548..820B} emphasised, in the
case of older and cooler SNRs, only $T_s$ can be determined from spatially
integrated X-ray observations with modest spectral resolution. We indeed found
little or no variations in the best-fit parameters and the $\chi ^2$ when
constraining $\beta = T_{es} / T_s$ between 0 (by taking $T_{es} = 0.01$ keV,
the minimal value in the Sedov model implemented in XSPEC) and 1. As a
consequence, we constrained the mean shock and postshock electron temperature to
be the same. This is a reasonable assumption, since old remnants should be close
to ion-electron temperature equilibrium.

The vpshock model parameters are the shock temperature and the upper limit to
the linear distribution of ionisation timescales $\tau _{up}$. This model has
been shown to approximate the Sedov model better than the commonly used
non-equilibrium ionisation (NEI) model \citep{2001ApJ...548..820B}. There are
deviations between vpshock and vsedov models for low-temperature shocks, but
these occur predominantly above 2~keV.

When fitted to the source region spectrum, the normalisations of the X-ray
background components were allowed to vary, but their \emph{ratios} were
constrained to be the same as in the background region. We found 5 \% or smaller
variations between the normalisations of the background components of the two
regions (which are shown by the dashed lines in Fig.\,\ref{fig_xray_spectrum}).
Because the background spectrum was extracted from the same observation
(that is, at the same time period) and at a similar position and off-axis angle
as the source spectrum, the SPC contribution was not expected to vary much
\citep{2008A&A...478..575K}. The validity of this assumption was checked \emph{a
posteriori} by looking at our data above 3 keV. We therefore used the same SPC
parameters for the background and source spectra.

To account for the absorption of the source emission, we included two
photoelectric absorption components, one with a column density N$_{H\mathrm{\
Gal}}$ for the Galactic absorption and another one with N$_{H\mathrm{\ LMC}}$
for the LMC. Except for O and Fe, which were allowed to vary, the metal
abundances for the source emission models were fixed to the average metallicity
in the LMC \citep[\emph{i.e.}, half the solar values,][] {1992ApJ...384..508R},
because the observations were not deep enough to permit abundance measurements
and because high-resolution spectroscopic data were unavailable.

\subsubsection{Spectral results}
\label{data_xray_spectral_results}

We fitted the data between 0.2 keV and 7 keV. We extended the fit down to low
energies to constrain the parameters of the LHB component, which had a low
plasma temperature ($kT \la 0.1$~keV). The data above 2~keV, where the Galactic
components hardly contribute, were necessary to constrain the non-thermal
extragalactic emission and the SPC \citep{2010ApJS..188...46K}.

The quality of our data statistics was too low to place strong constraints on
the foreground hydrogen absorption column. The best fit value for N$_{H\mathrm{\
Gal}}$ was 5.3 $\times\ 10^{20}$ cm$^{-2}$ (using the APEC component), with a 90
\% confidence interval from 3 to 10 $\times\ 10^{20}$ cm$^{-2}$. We therefore
fixed it at 5.9 $\times\ 10^{20}$ cm$^{-2}$ \citep[based on the \ion{H}{i}
measurement of][] {1990ARA&A..28..215D}. We found that the best-fit intrinsic
LMC column density value tended to 0, with a 90 \% confidence upper limit of 3.9
$\times\ 10^{20}$ cm$^{-2}$ (using the APEC component), and then fixed
N$_{H\mathrm{\ LMC}}$ to 0. We note that even though the best-fit temperature of
the Local Hot Bubble we derived (85 eV) agrees well with the results of
\citet{2008ApJ...676..335H}, the errors are large because this component
contributes only to a small number of energy bins. The significance of the LHB
component was less than 10\,\% (using a standard F-test), and we removed this
component from our final analysis. The power-law component was also faint but
more than 99.99\,\% significant.

We achieved good fits and obtained significant constraints on the source
parameters. The reduced $\chi^2$ were between 0.91 and 0.92. The plasma
temperatures ($kT$ between 0.2 keV and 0.3 keV) are consistent for all models.
They are similar to temperatures found in other extensive SNRs
\citep[\emph{e.g.}][] {2004ApJ...613..948W,2010ApJ...725.2281K}. The unabsorbed
X-ray luminosity of the Sedov model is $1.43 \times 10^{35}$ erg\,s$^{-1}$ in
the range 0.2\,--\,5 keV, whilst the other models yield similar values. More
than 90\,\% of the energy is released below 0.9 keV.

The best-fit values with 90\,\% confidence levels (CL) errors are listed in
Table \ref{table_spectral_results}. The spectrum fitted by the best-fit Sedov
model is shown in Fig.\,\ref{fig_xray_spectrum}. The ionisation timescales were
large (more than $10^{12}$ s cm$^{-3}$), which indicates quasi-equilibrium. In
this regime, $kT$ and $\tau$ are degenerate, because the spectra hardly change
when increasing $kT$ and decreasing $\tau$. This effect is shown in
Fig\,\ref{fig_xray_spectrum}. However, the emission measure (EM), which is a
function of the volume of emitting plasma and densities, \emph{does not} depend
on the model used, provided the column density is the same. With the help of
the 99\,\% CL range of EM obtained using the APEC model (16.0 -- 25.4 $\times
10^{57}$ cm$^{-3}$), which does not have the $kT$\,--\,$\tau$ degeneracy
problem, we obtained additional constraints on $kT$ and $\tau$.

The O and Fe abundances are (about 0.2 dex) lower than those in
\citet{1992ApJ...384..508R} but consistent with the results found by
\citet{1998ApJ...505..732H} in other LMC SNRs. The abundances found for \dem\
match well those reported in the nearby (13\arcmin\ or $\sim$190 pc in
projection) LMC SB N\,51D \citep{2010ApJ...715..412Y}.

\begin{figure*}[t!]
	\centering
\includegraphics
[bb= 52 111 514 635, clip, angle=-90,
width=0.3765\hsize]{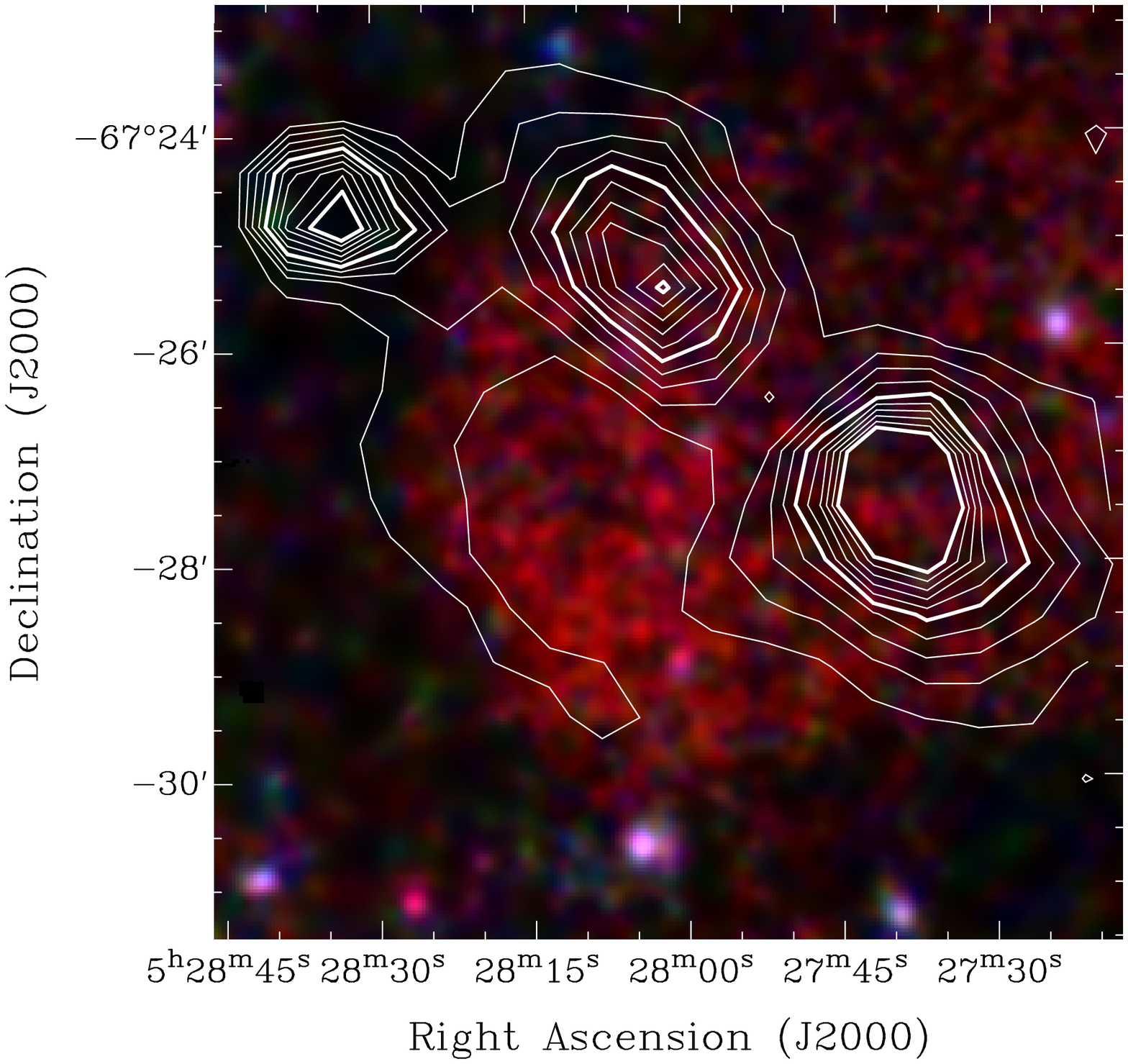}
\includegraphics
[bb= 52 208 567 635, clip, angle=-90,
width=0.3068\hsize]{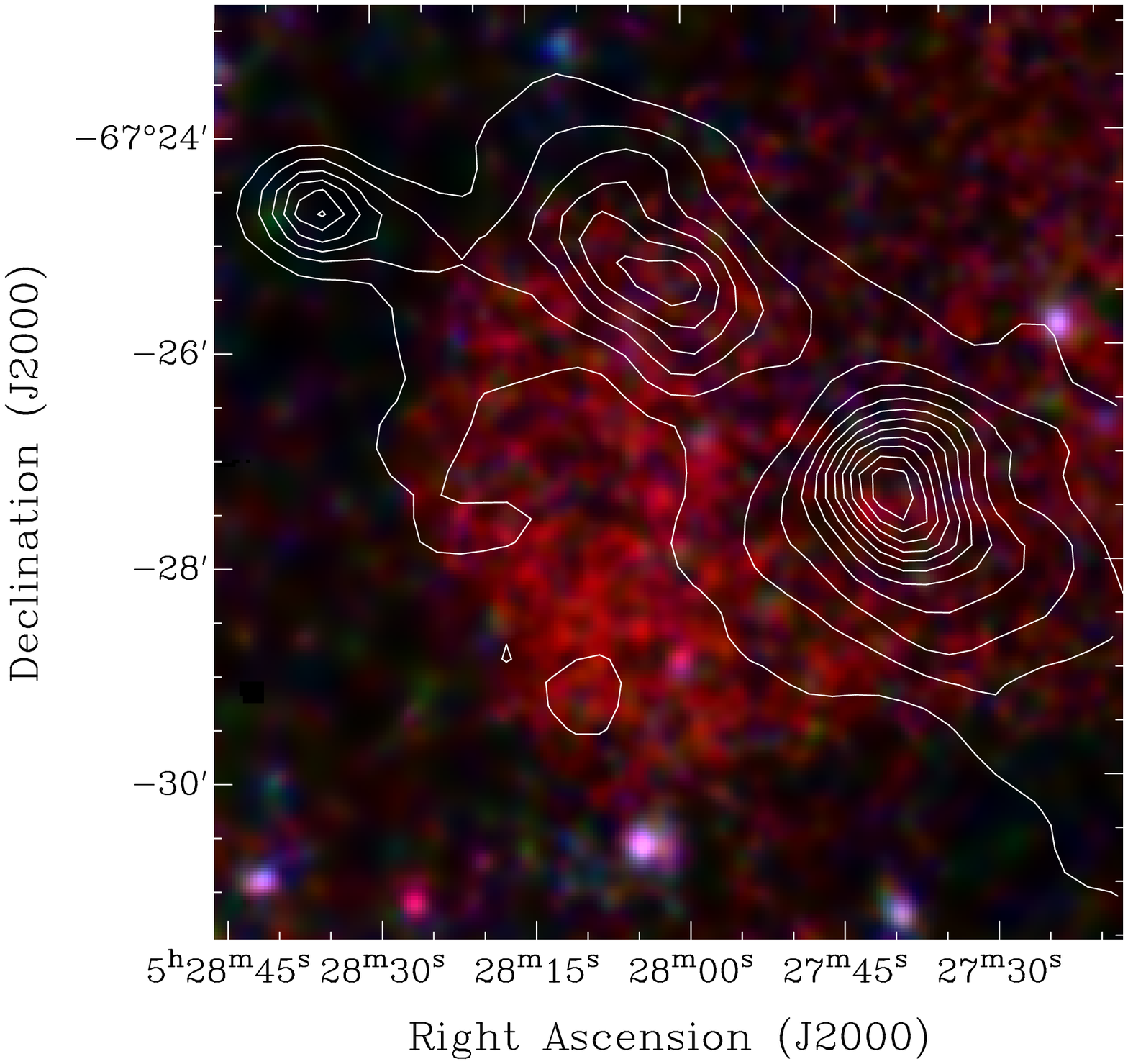}
\includegraphics
[bb= 52 208 514 635, clip, angle=-90,
width=0.3068\hsize]{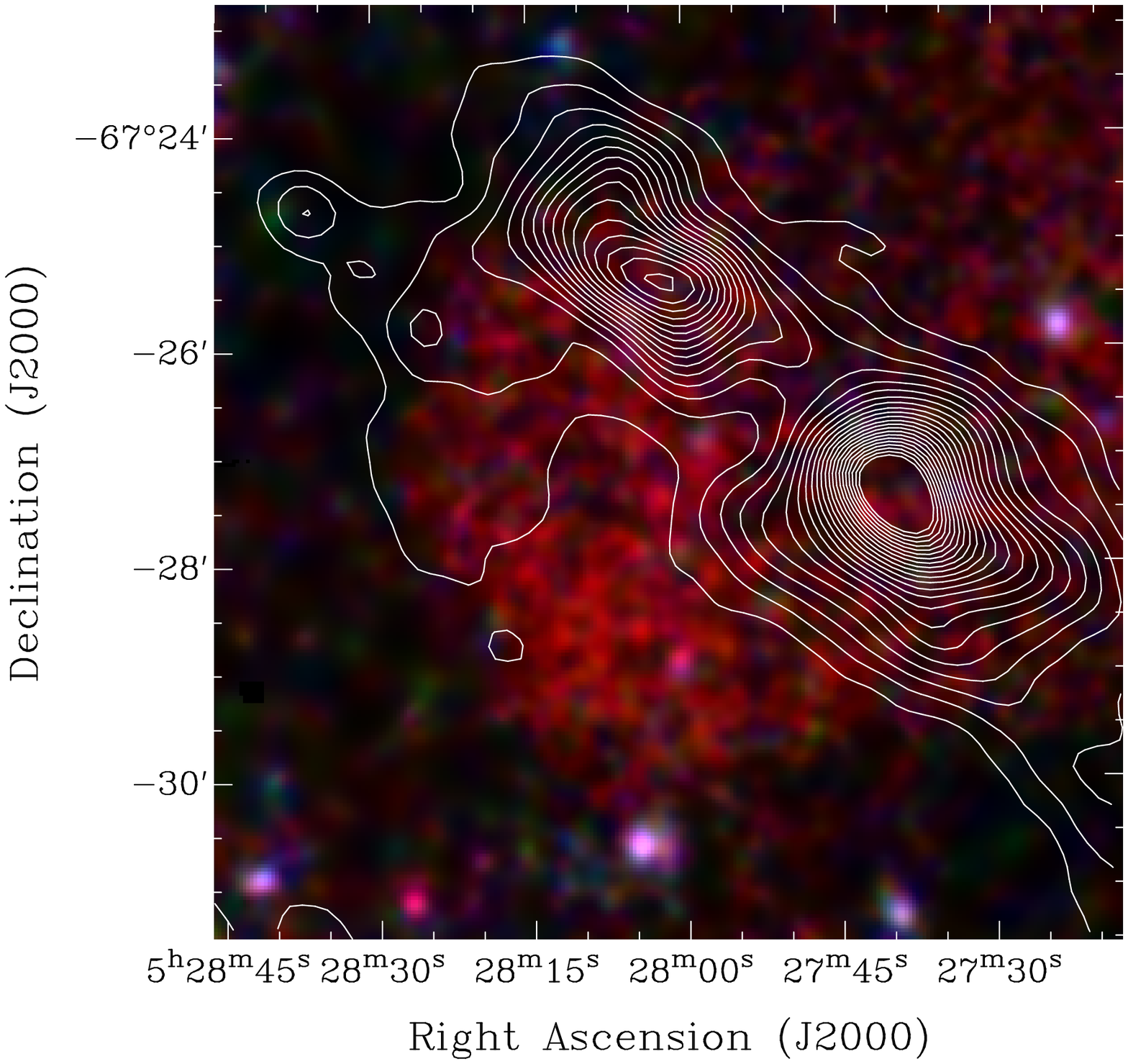}

\includegraphics[width=0.325\hsize]
{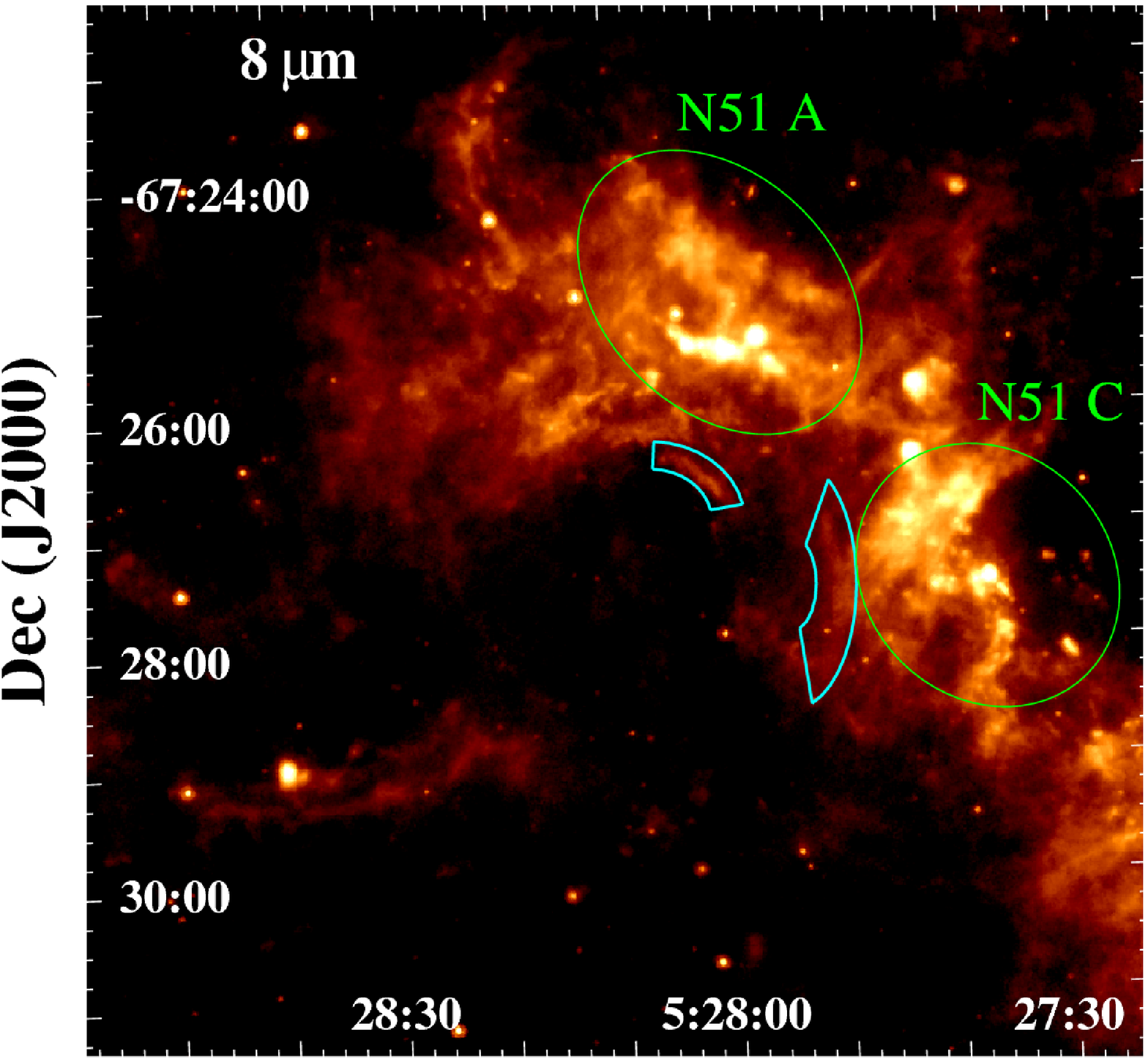}
\includegraphics[width=0.325\hsize]
{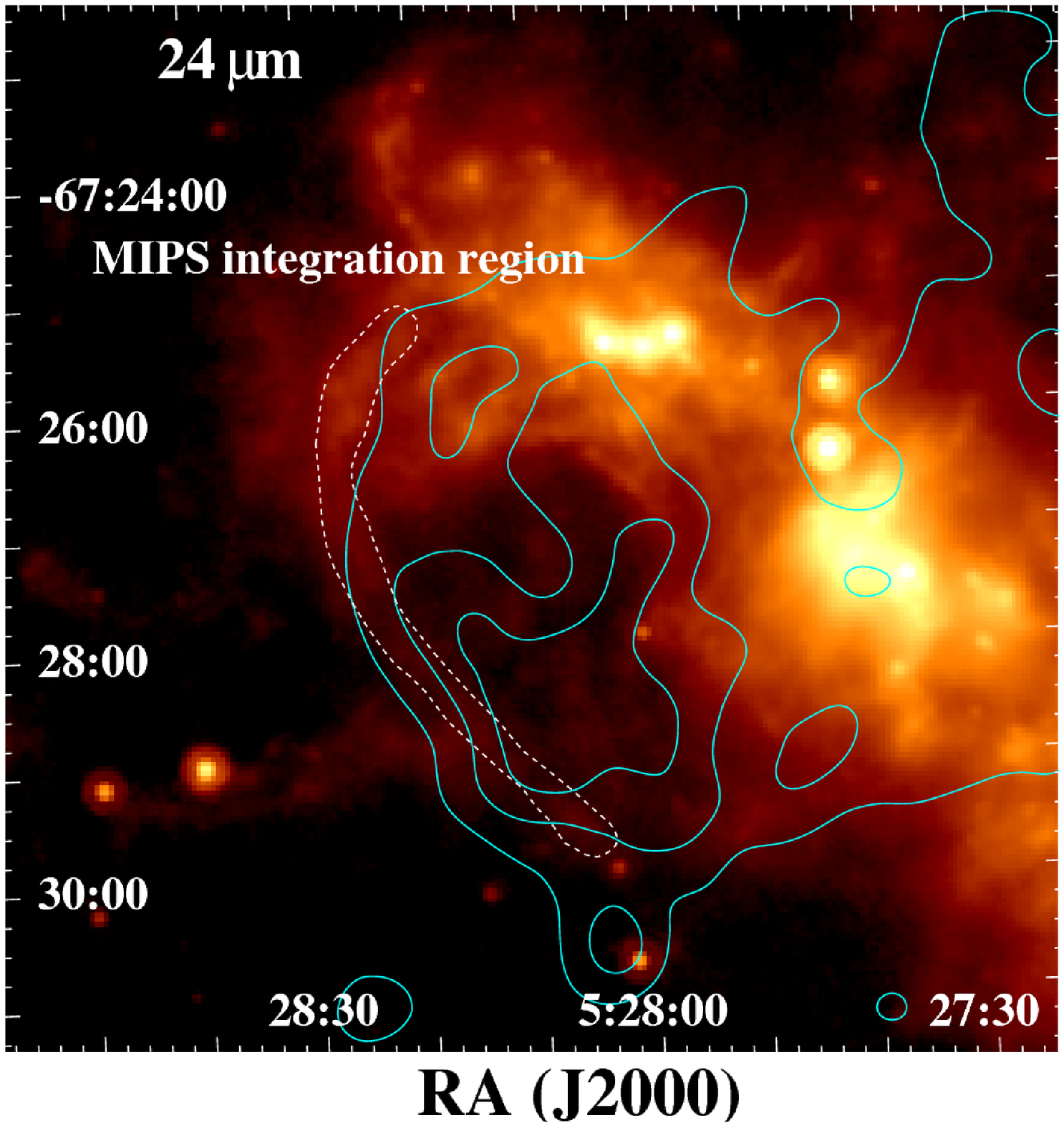}
\includegraphics[width=0.325\hsize]
{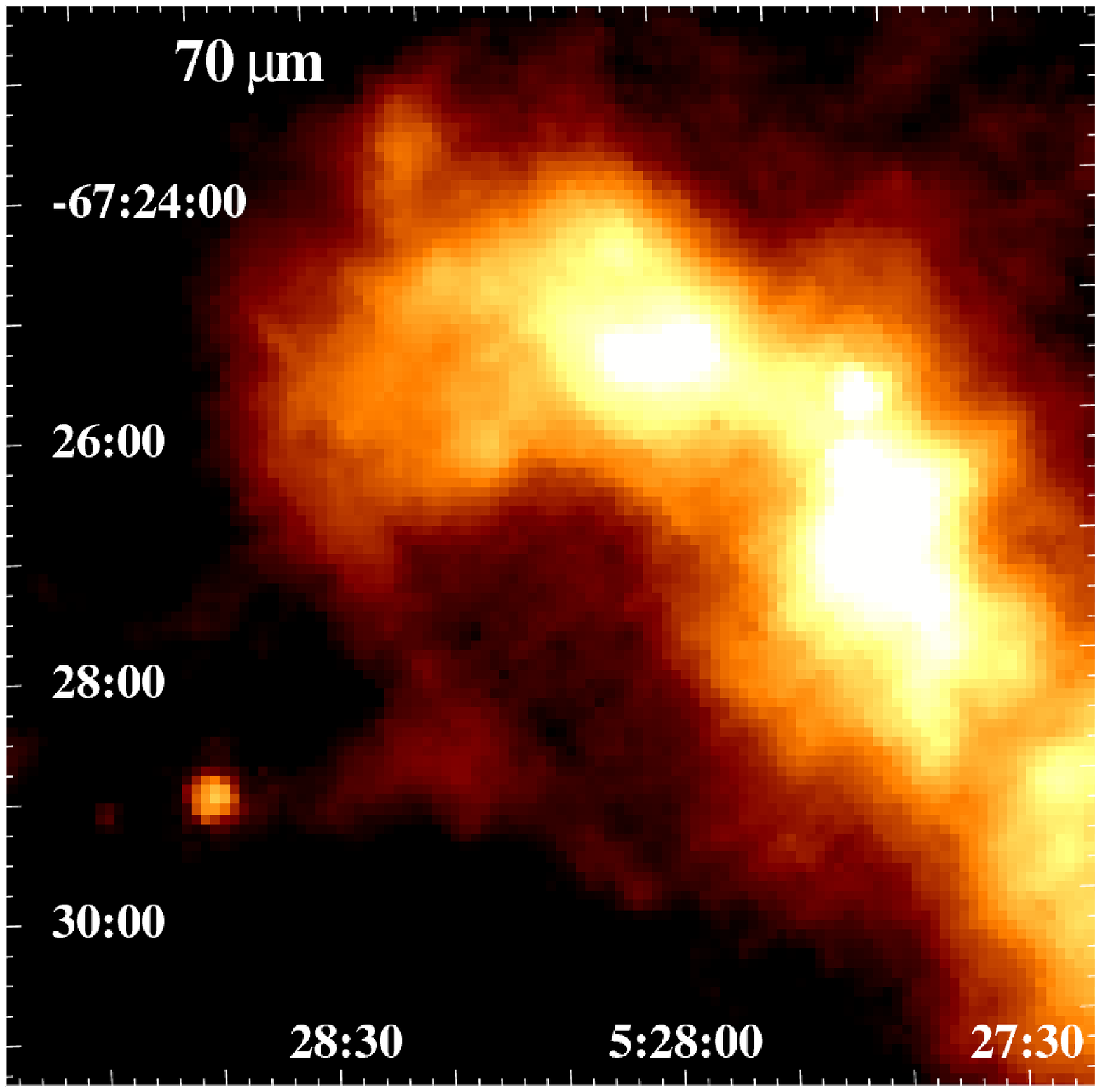}

\caption{
\emph{Top row\,:} Radio contours of \dem\ overlaid on the X-ray image.
\emph{Left\,:} 36~cm contours from 5 to 50$\sigma$ with 5$\sigma$ steps
($\sigma$=0.4~mJy/beam).
\emph{Middle\,:} 20~cm contours from 3 to 23$\sigma$ with 2$\sigma$ steps
($\sigma$=1.3~mJy/beam).
\emph{Right\,:} 6~cm contours from 3 to 200$\sigma$ with 9$\sigma$ steps
($\sigma$=0.1~mJy/beam).
Beam sizes are 40\arcsec\,$\times$\,40\arcsec\ for the 36 and 20~cm images, and
41.8\arcsec\,$\times$\,28.5\arcsec\ at 6~cm.
Note that the portion of the sky shown is smaller than in
Fig.\,\ref{fig_rgb_image}.\newline
\emph{Bottow row\,:} \emph{Spitzer} images of \dem\ at 8, 24, and 70 $\mu$m
(from  left to right). All images show a similar portion of the sky as the radio
images and are displayed logarithmically. The green ellipses on the 8 $\mu$m
image show the positions of the two \ion{H}{ii} regions seen in the 36\,cm image
(top left), and the cyan arcs indicate the 8 $\mu$m emission possibly associated
with the SNR. The white dashed line shown in the 24 $\mu$m image marks the
region where we measured the flux densities at 24 $\mu$m and 70 $\mu$m
(Sect.\,\ref{data_ir}). Soft X-ray contours are overlaid in cyan.
}
\label{fig_radio_ir}
\end{figure*}

\subsection{Radio}
\label{data_radio}

\subsubsection{Morphology}
\label{data_radio_morphology}

To assess the morphology of the source, we overlaid the radio contours on the
\xmm\ image. Weak, extended ring-like emission correlates with the
eastern side of the X-ray remnant and is most prominent, as expected, at 36~cm,
but only marginally detected at higher frequencies (Fig.\,\ref{fig_radio_ir},
top). It is difficult to classify the morphology of this SNR at radio
wavelengths because it lies in a crowded field. The surrounding radio emission
is dominated in the north by LHA 120-N 51A, which is classified as an
\ion{H}{ii} region \citep{1998A&AS..130..421F} and also correlates with the
small molecular cloud [FKM2008] LMC N J0528$-$6726 \citep{2008ApJS..178...56F},
and in the west by the \ion{H}{ii} region LHA 120-N 51C.

If we assumed that the analysable region of the 36~cm image
(Fig.\,\ref{fig_radio_ir}, top left) is typical of the rest of the remnant's
structure, the SNR would have a typical ring morphology. Nevertheless, with the
present resolution, one cannot easily estimate the total flux density of this
SNR at any radio frequency. However, we note the steep drop (across the eastern
side of the ring) in flux density at higher frequencies, which results in a
nearly completely dissipated remnant as seen at 6~cm (Fig.\,\ref{fig_radio_ir},
top right).

\subsubsection{Radio-continuum spectral energy distribution}
\label{data_radio_spectrum}
\begin{figure}[t]
	\centering
	\includegraphics[bb = 0 0 462 418,clip,width=\hsize]
{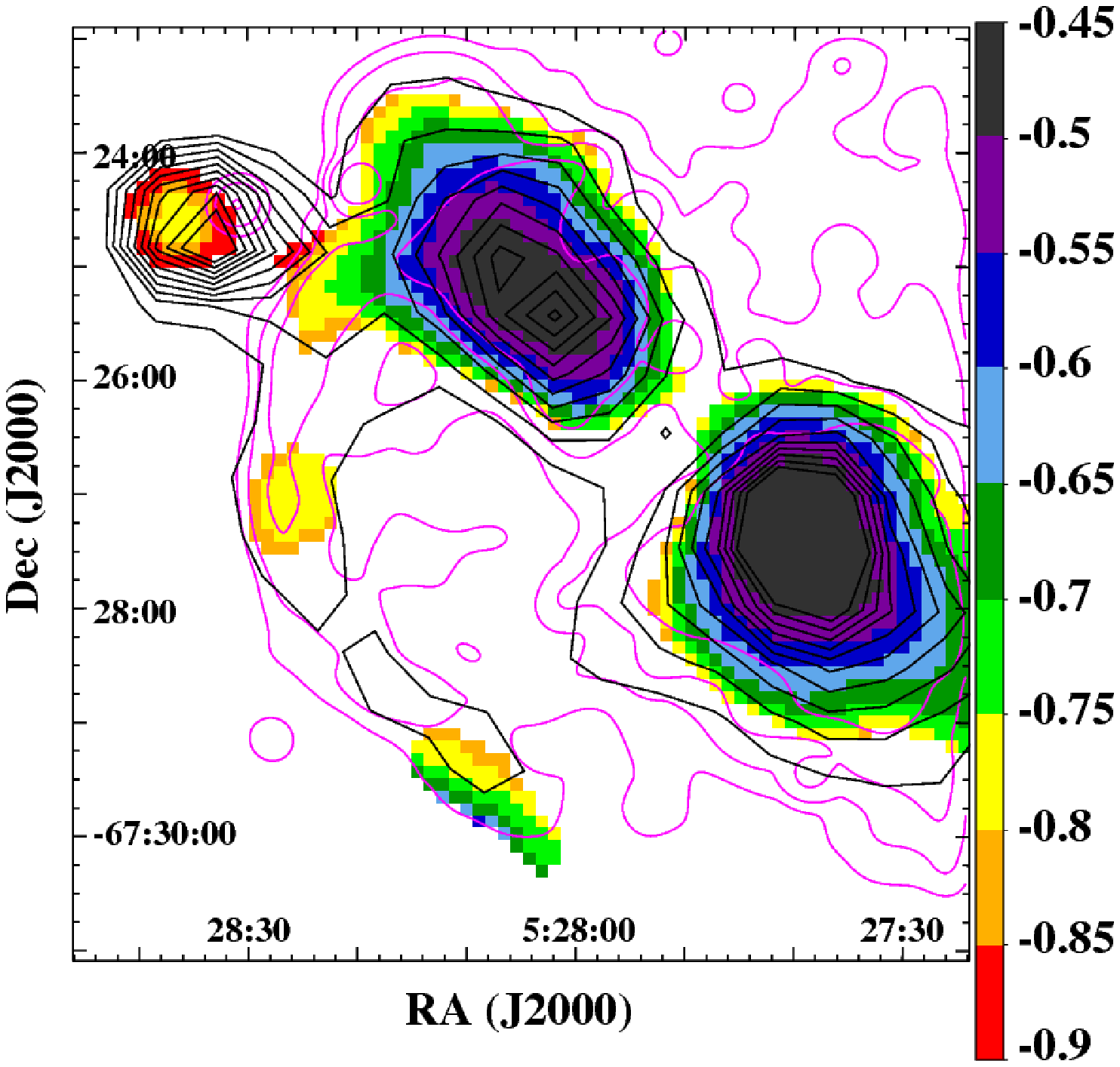}
	\caption{Spectral index map of \dem\ between wavelength of 36, 20, and 6~cm,
covering the same field as Fig\,\ref{fig_radio_ir}. The sidebar gives the
spectral index $\alpha$, as defined in the introduction. The 36~cm contours
(black) are overlaid, with the same levels as in Fig.\,\ref{fig_radio_ir}. The
H$\alpha$ structures (Figs.\,\ref{fig_rgb_N51} \& \ref{fig_rgb_image}) are
sketched by the magenta contours.}
\label{radio_spectral_map}
\end{figure}

We were unable to compile a global spectral index for the remnant because a
large portion of \dem\ cannot be analysed at radio wavelengths (as described in
Sect.\,\ref{data_radio_morphology}). However, a spectral index map
(Fig.\,\ref{radio_spectral_map}) shows the change in flux density from 36~cm to
6~cm. The map was formed by reprocessing all observations to a common $u-v$
range, and then fitting $S\propto \nu\, ^\alpha$ pixel by pixel using all three
images simultaneously. The areas of the SNR that are uncontaminated by strong
sources have spectral indices between $- 0.7$ and $- 0.9$, which is steeper but
close to the typical SNR radio-continuum spectral index of $\alpha \sim -0.5$.
We note that uncertainties in the determination of the background emission are
likely to cause a bias toward steeper spectral indices. We also point out that
the bright point source seen in the north-east (mainly at 36~cm) is most likely
a background galaxy or an active galactic nucleus (AGN).

\subsection{Infrared flux measurement}
\label{data_ir}
The IR data suffer from the same crowding issues as the radio-continuum data.
The IR emission in the SNR region (see Figure \ref{fig_radio_ir}, bottom) is
dominated by the two \ion{H}{ii} regions seen in radio, whose positions are
shown in the 8 $\mu$m images (Fig\,\ref{fig_radio_ir}, bottom left). However, at
24 $\mu$m, an arc of shell-like emission is seen in the eastern and
south-eastern regions of the remnant (outlined in Fig\,\ref{fig_radio_ir},
bottom middle) at the same position of the 36 cm emission. We used the optical
and X-ray emission contours to constrain the region at 24 $\mu$m that can be
truly associated with the SNR and found that this arc tightly follows the
H$\alpha$ and X-ray morphologies. We integrated the 24 $\mu$m surface brightness
in this region (in white in Fig\,\ref{fig_radio_ir}, bottom middle) and found a
flux density of $F_{24} = 660$ mJy.

To calibrate our method of flux density measurement and estimate the
uncertainties, we derived the 24 $\mu$m flux densities of the LMC SNRs N132D,
N23, N49B, B0453--68.5, and DEM L71, and compared them to the values published
in \citet{2006ApJ...642L.141B} and \citet{2006ApJ...652L..33W}. We were able to
reproduce these authors' values, but with rather large error ranges ($\sim$
30\,\%), chiefly because of uncertainties in the definition of the integration
region. The two aforementioned studies integrated the flux density only in
limited areas of the SNRs, and the integration regions are not explicitly
defined in their papers. In the case of \dem\ it is also difficult to define the
area of IR emission from the SNR only, so we believe these 30\,\% error ranges
are a reasonable estimate of the error in the flux density measurement. The
systematic uncertainties in the flux calibration of the \emph{Spitzer} images
are small in comparison and can be neglected. In particular, given that the
thickness of our region in the plane of the sky is 20\arcsec--25\arcsec, only a
small aperture correction would be needed (at least at 24 $\mu$m).

The 70 $\mu$m image (Fig\,\ref{fig_radio_ir}, bottom right) shows that \dem\ has
the same morphology as at 24 $\mu$m, but with lower resolution, hence the
confusion is even higher. Simply using the same region as for $F_{24}$, we found
a flux density of $F _{70} = 3.4$ Jy, with similarly large errors. We discuss
the origin of the IR emission in Sect.\,\ref{discussion_ir_emission}.

In the IRAC wavebands, no significant shell-like emission is detected. We
tentatively identified two arcs at 8 $\mu$m  (marked in cyan on
Fig\,\ref{fig_radio_ir}, bottom left) that could originate from the interaction
of the shock with higher densities towards the \ion{H}{ii} regions. The two arcs
are also present at 5.8 $\mu$m (not shown) but neither at 4.5 $\mu$m nor 3.6
$\mu$m, where only point sources are seen.

%
\section{Discussion}
\label{discussion}

Since the source exhibits all the classical SNR signatures, we can confirm \dem\
as a \emph{new supernova remnant}. This brings the total number of SNRs in the
LMC to 56, using the list of 54 remnants assembled by
\citet{2010MNRAS.407.1301B} and the new SNR identified by
\citet{2012A&A...539A..15G}. For the naming of SNRs in the Magellanic Clouds, we
advocate the use of the acronym ``MCSNR'', which was pre-registered to the
International Astronomical Union by R. Williams et al., who maintain the
Magellanic Cloud Supernova Remnants online
database\,\footnote{\url{http://www.mcsnr.org/Default.aspx}}. This should
ensure that there us a more consistent and general naming system for future
studies using the whole sample of SNRs in the LMC. On the basis of the J2000
X-ray position, \dem\ would thus be called \snr.

In the following sections, we take advantage of the multi-wavelength
observations of the remnant and discuss the origin of the IR emission
(Sect.\,\ref{discussion_ir_emission}), derive some physical properties of the
remnant (Sect.\,\ref{discussion_xray_properties}), compare the morphology at all
observed wavelengths and discuss the environment in which the SN exploded
(Sect.\,\ref{discussion_multi_frequency}), and analyse the star formation
activity around the SNR (Sect.\,\ref{discussion_star_formation}).

\subsection{Origin of the IR emission}
\label{discussion_ir_emission}
Supernova remnants emit IR light chiefly in forbidden lines,
rotational/vibrational lines of molecular hydrogen, emission in polycyclic
aromatic hydrocarbon (PAH) bands, and thermal continuum emission from dust
collisionally heated by shock waves \citep[\emph{e.g.}][]{2007PASJ...59S.455K}
and/or by stellar-radiation reprocessing. Infrared synchrotron emission is only
expected in pulsar wind nebulae, for instance in the Crab
\citep{2006AJ....132.1610T}. Polycyclic aromatic hydrocarbons are thought to be
destroyed by shocks with velocities higher than 100 km\,s$^{-1}$ and should not
survive for more than a thousand years in a tenuous hot gas
\citep{2010A&A...510A..37M,2010A&A...510A..36M}. However, PAH features have been
detected in Galactic SNRs \citep{2011ApJ...742....7A}, where shock velocities
are rather low owing to interactions with a molecular cloud environment, and
even in the strong shocks of the young LMC remnant N132D
\citep{2006ApJ...653..267T}.

No significant emission from \dem\ was detected in the IRAC wavebands (which
have been chosen to include the main PAH features), with the possible exception
of the two 8 and 5.8 $\mu$m arcs (Fig.\,\ref{fig_radio_ir}, bottom left) in the
direction of the neighbouring \ion{H}{ii} regions (in the north and west). This
means that PAHs have been efficiently destroyed. The absence of IR spectroscopic
observations precludes further interpretation.

\begin{figure*}[t]
	\centering
	\includegraphics[bb= 70 50 285 396,clip,height=11.2cm]{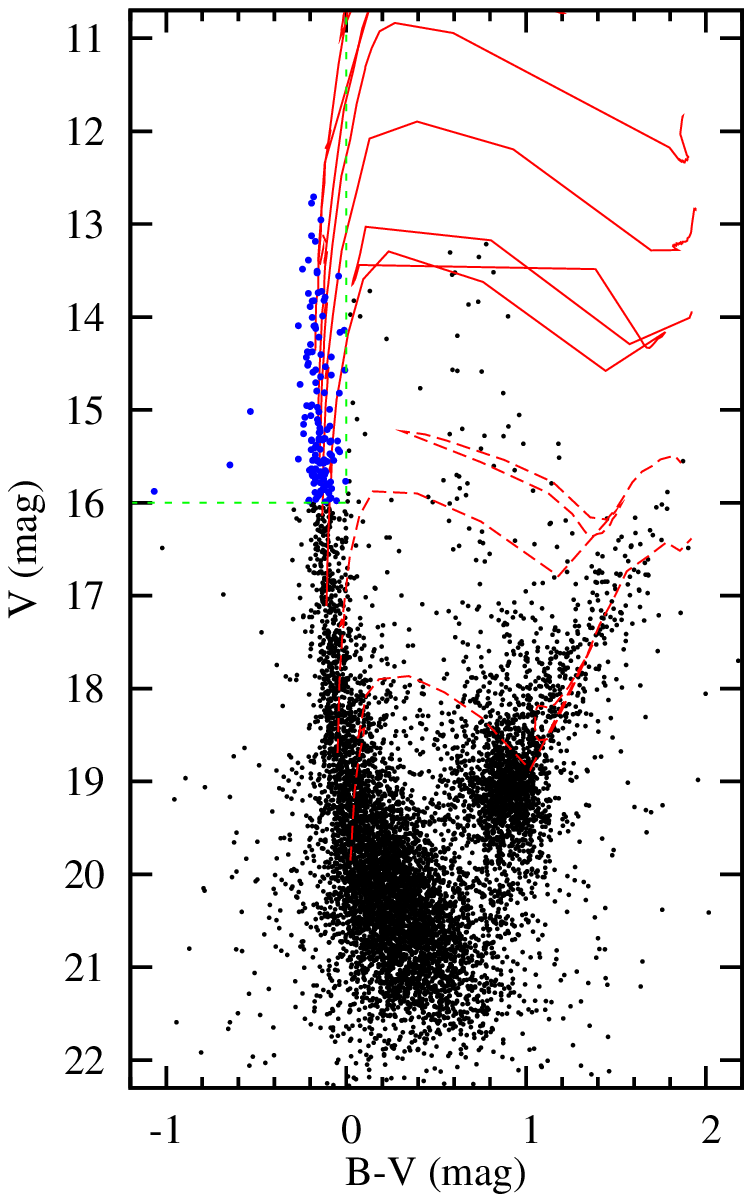}
\includegraphics[height=11.2cm]{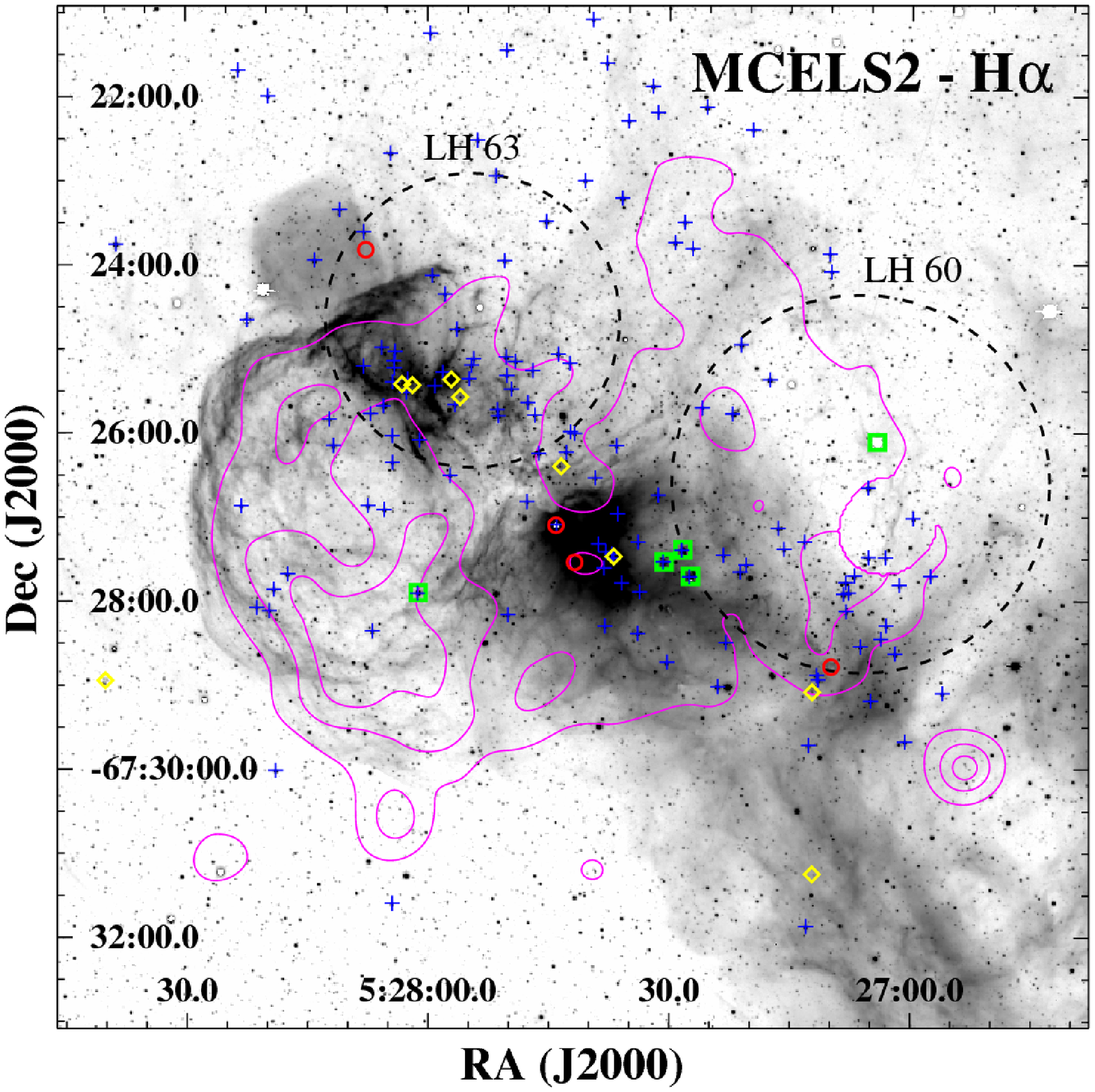}
	\caption{\emph{Left}\,: Colour-magnitude diagram (CMD) of the MCPS stars
\citep{2004AJ....128.1606Z} within 100 pc ($\sim$ 6.9\arcmin) of the central
position of \dem. Geneva stellar evolution tracks \citep{2001A&A...366..538L}
are shown as red lines, for metallicity of 0.4 Z$_{\sun}$ and initial masses of
3, 5 M$_{\sun}$ (dashed lines) and 10, 15, 20, 25, and 40 M$_{\sun}$ (solid
lines), from bottom to top. The green dashed line shows the criteria used to
identify the OB stars (V $<$ 16 and B$-$V $<$ 0). Stars satisfying these
criteria are shown as blue dots.\newline
\emph{Right}\,: MCELS2 H$\alpha$ image of the SNR, with the soft X-ray contours
in magenta. The blue plus signs show the positions of the OB candidates
identified in the CMD and green squares identify Sanduleak OB stars. The black
dashed circles encompass the nearby OB associations 60 and 63 from
\citet{1970AJ.....75..171L}. Positions of definite (yellow diamond) and probable
(red circle) YSOs from \citet{2009ApJS..184..172G} are also shown. }
\label{fig_environment}
\end{figure*}

The presence of H$\alpha$ emission shows that hydrogen is not in the molecular
phase, hence rotational/vibrational line contribution is negligible. The
emission in the 24 and 70 $\mu$m wavebands should then be dominated either by
dust or ionic forbidden lines. Ionic lines in the 24 $\mu$m filter bandpass
are [\ion{S}{i}] 25.2 $\mu$m, [\ion{Fe}{ii}] 24.50 and 25.99 $\mu$m,
[\ion{Fe}{iii}] 22.95 $\mu$m, and [\ion{O}{iv}] 25.91 $\mu$m. [\ion{S}{i}]
emission is not expected because of the prominent [\ion{S}{ii}] optical
emission, showing that \element[+]{S} is the primary ionisation stage of
sulphur. The morphological similarities between the MIR and X-ray emission lead
us to the interpretation that we mainly observe the thermal continuum of dust.
The correlation with 70 $\mu$m supports this scenario, and the 70-to-24 $\mu$m
ratio ($\sim 5.2$) is consistent with a dust temperature of 50--80 K
\citep{2006ApJ...652L..33W}. We note, however, that in the northern part of the
arc of 24 $\mu$m emission (encompassed by the white dashed line in
Fig.\,\ref{fig_radio_ir}, bottom middle), the MIR emission is slightly ahead of
the shock (delineated by the X-ray emission), whereas it correlates tightly with
the shock in the rest of the arc. This morphology and the presence of the OB
association \object{LH\,63} (see Fig.\,\ref{fig_environment}, right) indicates
that stellar radiation dominates the heating of the dust in the north. In the
southern part, shock waves could play a more significant role in heating the
dust.

The lack of spectroscopic data prevent us from establishing the precise
contribution of dust \emph{vs.} O and Fe lines. Because of these limitations and
the confusion with the background, and because only part of the SNR is detected
at IR wavelengths, we did not attempt to derive a dust mass. Consequently, no
dust-to-gas ratio (using the swept-up gas mass estimate from X-ray observations)
and dust destruction percentage can be given.

\subsection {Properties of DEM L205 derived from the X-ray observations}
\label{discussion_xray_properties}

From the X-ray spectral analysis, we can derive several physical properties of
the remnant\,: electron and hydrogen densities $n_\mathrm{e}$ and
$n_\mathrm{H}$, dynamical and ionisation ages $t_{\mathrm{dyn}}$ and
$t_{\mathrm{i}}$, swept-up mass $M$, and initial explosion energy $E_0$. We used
a system of equations adapted from \citet[]{2004A&A...421.1031V}, given by
\begin{equation}
\label{eq_ne}
	n_{\mathrm{e}} = \frac{1}{f} \sqrt{r_\mathrm{e} \, \frac{\mathrm{EM}}{V}}
	\quad \left(\mathrm{cm}^{-3}\right)
\end{equation}

\begin{equation}
\label{eq_nh}
	n_{\mathrm{H}} = n_{\mathrm{e}} / r_\mathrm{e}
	\quad \left(\mathrm{cm}^{-3}\right)
\end{equation}

\begin{equation}
\label{eq_tdyn}
	t_{\mathrm{dyn}} = 1.3 \times 10^{-16} \, \frac{R}{\sqrt{kT_s}} 
	\quad (\mathrm{yr})
\end{equation}

\begin{equation}
\label{eq_ti}
	t_{\mathrm{i}} = 3.17 \times 10^{-8} \, \frac{\tau}{n_\mathrm{e}} 
	\quad \left(\mathrm{yr}\right)
\end{equation}

\begin{equation}
\label{eq_M}
	M = 5 \times 10^{-34}\, m_\mathrm{p}\, r_\mathrm{m}\, n_\mathrm{e}\, f^2\, V
	\quad \left(\mathrm{M_{\sun}}\right)
\end{equation}

\begin{equation}
\label{eq_E0}
	E_0 = 2.64 \times 10 ^{-8} \, kT_s \, R^3 \, n_\mathrm{H} 
	\quad \left(\mathrm{erg}\right),
\end{equation}
where EM is the emission measure ($= n_\mathrm{e} n_\mathrm{H} V$) in cm$^{-3}$,
$kT_s$ is the shock temperature in keV, and $\tau$ is the ionisation timescale
in s\,cm$^{-3}$. These parameters are determined by the spectral fitting. In
addition, $R$ is the radius of the X-ray remnant in cm (using the semi-major
axis of 39.5 pc, see Sect.\,\ref{data_xray_image}), $V$ is the volume ($4\pi /3
\times R^3$) assuming spherical symmetry (as discussed below), $m_\mathrm{p}$ is
the proton mass in g, $r_\mathrm{m}$ is the total number of baryons per hydrogen
atom ($= n_\mathrm{m} / n_\mathrm{H}$), and $r_\mathrm{e}$ is the number of
electrons per hydrogen atom ($= n_\mathrm{e} / n_\mathrm{H}$). Assuming a plasma
with 0.5 solar metal abundances, as done in the spectral fitting, we have
$r_\mathrm{m} \approx 1.40$ and $r_\mathrm{e} \approx 1.20$ (for full
ionisation). Finally, $f$ is a filling factor to correct for any departure from
spherical symmetry, as inferred from the X-ray morphology. $f$ is defined as
$\sqrt{V_t/V}$, where $V_t$ is the true X-ray emitting (ellipsoidal) volume.
Adopting the semi-minor axis of the X-ray emitting ellipse (32 pc) as the second
semi-principal axis, $f$ is in the range 0.81 -- 0.90, with the third
semi-principal axis being between 32 pc and 39.5 pc. The properties are listed
in Table\,\ref{table_properties}, using $f$ in this range and EM in the range
defined in Sect.\,\ref{data_xray_spectral_results}.

\begin{table}[ht]
\caption{Physical properties of \dem.}
\label{table_properties}
\centering
\begin{tabular}{c c c c c c}
\hline
\hline
\noalign{\smallskip}
$n_\mathrm{e}$&$n_\mathrm{H}$&$t_{\mathrm{dyn}}$&$M$&$E_0$ \\
\multicolumn{2}{c}{$\left(10^{-2}\ \mathrm{cm}^{-3}\right)$} &
$\left(10^{3}\ \mathrm{yr}\right)$ &
$\left(\mathrm{M_{\sun}}\right)$ &
$\left(10^{51}\ \mathrm{erg}\right)$\\
\noalign{\smallskip}
\hline
\noalign{\smallskip}
5.6 -- 7.8 & 4.7 -- 6.5 & 35$^{+2} _{-5}$& 400 -- 460 & 0.52 -- 0.77\\
\noalign{\smallskip}
\hline
\end{tabular}
\end{table}

The large amount (from 400 M$_{\sun}$ to 460 M$_{\sun}$) of swept-up gas
justifies \emph{a posteriori} that the SNR is indeed well-established in the
Sedov phase. Because the remnant is old, the plasma is close to or in
collisional ionisation equilibrium, as indicated by either the acceptable fit of
the APEC model or the large ionisation timescale $\tau$, for which only a lower
limit is found. Thus, the spectrum changes very slowly with time and $\tau$ is
no longer a sensitive age indicator \citep{2004A&A...421.1031V}. This explains
why $t_i$ is unrealistically long ($> 770$ kyr, from Eqs.\,\ref{eq_ne} and
\ref{eq_ti}) and unreliable.

\subsection{Multi-wavelength morphology}
\label{discussion_multi_frequency}
In Fig.\,\ref{fig_rgb_image}, we see an X-ray remnant with a slightly elongated
shape and a maximal extent of 79 pc. Therefore, \dem\ ranks amongst the largest
known in the LMC \citep[compared \emph{e.g.} to SNRs described in][]
{2009SerAJ.179...55C,2010ApJ...725.2281K,2012A&A...539A..15G}. The optical
emission-line images (Figs.\,\ref{fig_rgb_image} and \ref{fig_environment}) show
the shell-like structure of \object{DEM L205} coinciding with the boundary of
the X-ray emission from the remnant. \citet{2001ApJS..136..119D} classified the
shell as a SB, interpreting the morphology of \object{DEM L205} as a blister
blown by the OB association \object{LH 63} (see
Sect.\,\ref{discussion_star_formation} and Fig.\,\ref{fig_environment}, right).
They measured a mild expansion velocity of $\sim$ 70 km\,s$^{-1}$ for the
H$\alpha$ shell, which is typical of SBs. Supernova remnants exhibit higher
expansion velocities ($\gtrsim$ 100 km\,s$^{-1}$), although this is not a
necessary condition \citep{1997AJ....113.1815C}.

On the basis of the low densities ($< 0.1$ cm$^{-3}$) derived from the X-ray
spectral analysis, we conclude that the supernova exploded inside the blister,
producing the bright X-ray emission in the interior of the SB. The SNR shocks
reaching the inner edge of the bubble might then have produced non-thermal radio
emission, and the observed morphology at 36~cm is consistent with this picture. 

The remnant is located in a complex environment. In the north and west, we
detected two \ion{H}{ii} regions and a strip of dust and gas extending down
towards the south-west. The \ion{H}{ii} regions also show bright IR emission,
mainly from dust heated by stellar radiation, and bright (thermal)
radio-continuum emission (Fig.\,\ref{fig_radio_ir}).
The [\ion{S}{ii}]-to-H$\alpha$ ratio is higher in the south of the remnant,
indicating that the diffuse optical emission there is caused by the SNR shocks.
In addition, the lower ratio in the north and west parts of the remnant
is most likely due to photoionisation by the massive stars (bringing sulphur to
ionisation stages higher than \element[+]{S}) from the same OB associations
that power the \ion{H}{ii} regions and produced the SB in which the supernova
exploded. We therefore propose that the SNR and the \ion{H}{ii} regions are
physically connected.

Furthermore, whilst the X-ray surface brightness falls abruptly across the
eastern and southern boundaries of the remnant, much weaker emission is detected
in the north and west, right at the positions of the \ion{H}{ii} regions seen at
all other wavelengths (Figs.\,\ref{fig_rgb_image} and \ref{fig_radio_ir}). This
indicates that the remnant is \emph{behind} the \ion{H}{ii} regions. The
absorption column density is higher in the north and west, suppressing the X-ray
emission and giving rise to the observed asymmetrical, irregular shape in these
regions. The ellipse defined in Sect.\,\ref{data_xray_image} is probably an
oversimplification of the actual morphology of the X-ray emitting region. The
remnant may have a more spherical shape, with some parts masked by the
\ion{H}{ii} regions.

We also detected soft and faint diffuse X-ray emission on the other side of the
dust\,/\,gas strip. The diffuse X-ray emission is enclosed by very sharp and
faint H$\alpha$ filaments (Figs.\,\ref{fig_rgb_image} and
\ref{fig_environment}). The presence of the OB association (\object{LH 60})
suggests that we observed another stellar-wind-blown SB in which a SN had
exploded. The faintness of the X-ray and optical emission precludes further
analysis. We note here that the [\ion{S}{ii}]-to-H$\alpha$ ratio is $< 0.4$\,.
However, it cannot be used in that case because of the presence of massive stars
from the OB association.

\subsection{Past and present star formation activity around the SNR}
\label{discussion_star_formation}
The star formation history (SFH) and high-mass-star content of the local
environment of \dem\ can help us to determine whether the remnant's supernova
progenitor type is either thermonuclear (type Ia) or core-collapse (CC). The
latter originates from massive stars that rarely form in isolation. Therefore,
the combination of a recent peak in the star formation rate and the presence of
many early-type stars is expected in the case of CC SNe, whereas the contrary
would be more consistent with a type Ia SN.

To investigate the star content around the remnant, we used the Magellanic
Clouds Photometric Survey (MCPS) catalogue of \citet{2004AJ....128.1606Z} and
constructed the colour-magnitude diagram (CMD) of the $\sim$20\,000 stars
lying within 100 pc (6.9\arcmin) of the remnant's centre. The CMD
(Fig.\,\ref{fig_environment}, left) shows a prominent upper main-sequence
branch. We added stellar evolutionary tracks of \citet{2001A&A...366..538L},
for Z $=$ 0.4 Z$_{\sun}$ and initial masses from 3 M$_{\sun}$ to 40 M$_{\sun}$,
assuming a distance modulus of 18.49 and extinction $A_V = 0.5$ \citep[the
average extinction for ``hot'' stars,][]{2004AJ....128.1606Z}. We used the
criteria of V $< 16$ and B$-$V $<$ 0 to identify OB stars, and found 142 of them
in our sample (shown in a H$\alpha$ image in Fig.\,\ref{fig_environment},
right). Using V $<$ 15 or 14 instead of 16 would give 86 and 20 stars,
respectively.

We also looked for nearby OB associations in \citet{1970AJ.....75..171L} and OB
stars in the catalogue of \citet{1970CoTol..89.....S}. Contamination by Galactic
stars was monitored by performing a cross-correlation with Tycho-2 stars
\citep{2000A&A...355L..27H}. Five Sanduleak stars are in this region, four of
them having a match in the MCPS catalogue, with our selection criteria. The
``missed'' Sanduleak star is a VV Cepheid (a binary with a red component),
which thus possibly explains why our criteria were not satisfied. Two OB
associations (LH 60 and 63) lie close to the remnant ($\sim$ 6\arcmin\ and
3\arcmin, respectively), and their extent indeed contain many OB stars from the
MCPS catalogue.

\citet{2008PASA...25..116H} performed a spatially resolved analysis of the SFH
of the ``Constellation III'' region, and \dem\ was included in their study
(the ``E00'' cell in their Fig.\,2). They identified that a very strong peak in
the star formation rate occurred in the region of the remnant 10 Myr ago, and
that little star formation activity had occurred prior to this burst.

The rich content of high-mass stars and the recent peak in SFH around the
remnant strongly suggest that a core-collapse supernova has formed \dem. It is
however impossible to completely rule out a type Ia event. Considering at face
value that most of the stars were formed in the SFR peak 10~Myr ago, we
estimated a lower limit for the mass of the SN progenitor of
20\,$\mathrm{M_{\sun}}$, because less massive stars have a lifetime longer than
10 Myr \citep{1994A&AS..103...97M}. We cannot estimate an upper limit, because
the progenitor might have formed more recently (the region is still actively
forming stars, see below).

We searched for nearby young stellar objects (YSOs) to assess the possibility of
SNR-triggered star formation, as in \citet{2010AJ....140..584D}. Using the YSOs
from the catalogue of \citet{2009ApJS..184..172G}, we report an SNR--molecular
cloud--YSOs association around \dem\,: the positions of young stars are
shown in our H$\alpha$ image (Fig.\,\ref{fig_environment}, right). Four YSOs lie
in the \ion{H}{ii} region/molecular cloud in the north, and are closely aligned
with the X-ray emission rim. In addition, four YSOs lie in the western
\ion{H}{ii} region, significantly beyond the remnant's emission but correlated
with the diffuse X-ray emission from the SB around \object{LH 60} (see
Sect.\,\ref{discussion_multi_frequency}). Two additional YSOs are aligned with
the south-western edge of the SB.

Given the contraction timescale for the intermediate to massive YSOs
\citep[$10^6$ yr to $10^5$ yr,][]{1996A&A...307..829B}, the shocks from the
remnant cannot have triggered the formation of the YSOs already present. These
YSOs are more likely to have formed by interactions with winds and ionisation
fronts from the local massive stars, as illustrated by the alignment of young
stars along the rim of the adjacent SB. The remnant will be able to trigger star
formation in the future, when the shocks have slowed down to below 45
km\,s$^{-1}$ \citep{1998ApJ...508..291V}. By this time, however, the
neighbouring massive stars will also have triggered further star formation. It
is therefore difficult to assess the exact triggering agent of star formation,
as \citet{2010AJ....140..584D} pointed out, in particular in such a complex
environment.

%
\section{Conclusions}
\label{conclusions}
The first observation of our LMC survey with \xmm\ included the SNR
candidate \dem\ in the field of view. In combination with unpublished
radio-continuum data and archival optical and IR observations, we have found all
classical SNR signatures, namely\,:
\begin{itemize}
	\item extended X-ray emission
	\item optical emission with a shell-like morphology and an enhanced
	[\ion{S}{ii}]-to-H$\alpha$ ratio
	\item non-thermal and extended radio-continuum emission.
\end{itemize}
The source is also detected in the IR where we predominantly observe thermal
emission from dust. We can therefore definitely confirm this object as a
supernova remnant. A core-collapse supernova origin is favored, in light of the
recent burst of star formation and the presence of many massive stars in the
close vicinity of the remnant. The SN exploded in a SB, thus expanding in a low
density medium. With a size of $\sim$ 79 \mbox{$\times$~64 pc}, \dem\ is one of
the largest SNR known in the LMC. Given the low plasma temperature ($kT \sim$
0.2\,--\,0.3 keV), we derived a dynamical age of about 35~kyr. Whilst completing
our survey, we can expect to find other similarly evolved remnants, thereby
refining the faint end of the size and luminosity distributions of SNRs in the
LMC.

\begin{acknowledgements}
The \xmm\ project is supported by the Bundesministerium f\"ur Wirtschaft und
Technologie\,/\,Deutsches Zentrum f\"ur Luft- und Raumfahrt (BMWi/DLR, FKZ 50 OX
0001) and the Max-Planck Society.
Cerro Tololo Inter-American Observatory (CTIO) is operated by the Association of
Universities for Research in Astronomy Inc. (AURA), under a cooperative
agreement with the National Science Foundation (NSF) as part of the National
Optical Astronomy Observatories (NOAO). We gratefully acknowledge the support of
CTIO and all the assistance which has been provided in upgrading the Curtis
Schmidt telescope.
The MCELS is funded through the support of the Dean B. McLaughlin fund at the
University of Michigan and through NSF grant 9540747.
The Australia Telescope Compact Array is part of the Australia Telescope which
is funded by the Commonwealth of Australia for operation as a National Facility
managed by CSIRO. We used the {\sc karma} software package developed by the
ATNF.
This research has made use of Aladin, SIMBAD and VizieR, operated at CDS,
Strasbourg, France.
Pierre Maggi thanks Philipp Lang for helpful discussions regarding IR data
analysis.
P.\,M. and R.\,S. acknowledge support from the BMWi/DLR grants FKZ 50 OR 1201
and FKZ OR 0907, respectively.
R.\,A.\,G. was partially supported by the NSF grant AST 08-07323.
\end{acknowledgements}



\begin{thebibliography}{59}
\expandafter\ifx\csname natexlab\endcsname\relax\def\natexlab#1{#1}\fi

\bibitem[{{Andersen} {et~al.}(2011){Andersen}, {Rho}, {Reach}, {Hewitt}, \&
  {Bernard}}]{2011ApJ...742....7A}
{Andersen}, M., {Rho}, J., {Reach}, W.~T., {Hewitt}, J.~W., \& {Bernard}, J.~P.
  2011, \apj, 742, 7

\bibitem[{{Arnaud}(1996)}]{1996ASPC..101...17A}
{Arnaud}, K.~A. 1996, in Astronomical Society of the Pacific Conference Series,
  Vol. 101, Astronomical Data Analysis Software and Systems V, ed.
  {G.~H.~Jacoby \& J.~Barnes}, 17

\bibitem[{{Badenes} {et~al.}(2010){Badenes}, {Maoz}, \&
  {Draine}}]{2010MNRAS.407.1301B}
{Badenes}, C., {Maoz}, D., \& {Draine}, B.~T. 2010, \mnras, 407, 1301

\bibitem[{{Balucinska-Church} \& {McCammon}(1992)}]{1992ApJ...400..699B}
{Balucinska-Church}, M. \& {McCammon}, D. 1992, \apj, 400, 699

\bibitem[{{Bernasconi} \& {Maeder}(1996)}]{1996A&A...307..829B}
{Bernasconi}, P.~A. \& {Maeder}, A. 1996, \aap, 307, 829

\bibitem[{{Borkowski} {et~al.}(2001){Borkowski}, {Lyerly}, \&
  {Reynolds}}]{2001ApJ...548..820B}
{Borkowski}, K.~J., {Lyerly}, W.~J., \& {Reynolds}, S.~P. 2001, \apj, 548, 820

\bibitem[{{Borkowski} {et~al.}(2006){Borkowski}, {Williams}, {Reynolds},
  {Blair}, {Ghavamian}, {Sankrit}, {Hendrick}, {Long}, {Raymond}, {Smith},
  {Points}, \& {Winkler}}]{2006ApJ...642L.141B}
{Borkowski}, K.~J., {Williams}, B.~J., {Reynolds}, S.~P., {et~al.} 2006, \apjl,
  642, L141

\bibitem[{{Cajko} {et~al.}(2009){Cajko}, {Crawford}, \&
  {Filipovic}}]{2009SerAJ.179...55C}
{Cajko}, K.~O., {Crawford}, E.~J., \& {Filipovic}, M.~D. 2009, Serbian
  Astronomical Journal, 179, 55

\bibitem[{{Chen} {et~al.}(1997){Chen}, {Fabian}, \&
  {Gendreau}}]{1997MNRAS.285..449C}
{Chen}, L.-W., {Fabian}, A.~C., \& {Gendreau}, K.~C. 1997, \mnras, 285, 449

\bibitem[{{Chu}(1997)}]{1997AJ....113.1815C}
{Chu}, Y.-H. 1997, \aj, 113, 1815

\bibitem[{{Davies} {et~al.}(1976){Davies}, {Elliott}, \&
  {Meaburn}}]{1976MmRAS..81...89D}
{Davies}, R.~D., {Elliott}, K.~H., \& {Meaburn}, J. 1976, \memras, 81, 89

\bibitem[{{de Horta} {et~al.}(2012){de Horta}, {Filipovi{\'c}}, {Bozzetto},
  {Maggi}, {Haberl}, {Crawford}, {Sasaki}, {Uro{\v s}evi{\'c}}, {Pietsch},
  {Gruendl}, {Dickel}, {Tothill}, {Chu}, {Payne}, \&
  {Collier}}]{2012A&A...540A..25D}
{de Horta}, A.~Y., {Filipovi{\'c}}, M.~D., {Bozzetto}, L.~M., {et~al.} 2012,
  \aap, 540, A25

\bibitem[{{Desai} {et~al.}(2010){Desai}, {Chu}, {Gruendl}, {Dluger}, {Katz},
  {Wong}, {Chen}, {Looney}, {Hughes}, {Muller}, {Ott}, \&
  {Pineda}}]{2010AJ....140..584D}
{Desai}, K.~M., {Chu}, Y.-H., {Gruendl}, R.~A., {et~al.} 2010, \aj, 140, 584

\bibitem[{{di Benedetto}(2008)}]{2008MNRAS.390.1762D}
{di Benedetto}, G.~P. 2008, \mnras, 390, 1762

\bibitem[{{Dickel} {et~al.}(2005){Dickel}, {McIntyre}, {Gruendl}, \&
  {Milne}}]{2005AJ....129..790D}
{Dickel}, J.~R., {McIntyre}, V.~J., {Gruendl}, R.~A., \& {Milne}, D.~K. 2005,
  \aj, 129, 790

\bibitem[{{Dickel} {et~al.}(2010){Dickel}, {McIntyre}, {Gruendl}, \&
  {Milne}}]{2010AJ....140.1567D}
{Dickel}, J.~R., {McIntyre}, V.~J., {Gruendl}, R.~A., \& {Milne}, D.~K. 2010,
  \aj, 140, 1567

\bibitem[{{Dickey} \& {Lockman}(1990)}]{1990ARA&A..28..215D}
{Dickey}, J.~M. \& {Lockman}, F.~J. 1990, \araa, 28, 215

\bibitem[{{Dunne} {et~al.}(2001){Dunne}, {Points}, \&
  {Chu}}]{2001ApJS..136..119D}
{Dunne}, B.~C., {Points}, S.~D., \& {Chu}, Y.-H. 2001, \apjs, 136, 119

\bibitem[{{Fazio} {et~al.}(2004){Fazio}, {Hora}, {Allen}, {Ashby}, {Barmby},
  {Deutsch}, {Huang}, {Kleiner}, {Marengo}, {Megeath}, {Melnick}, {Pahre},
  {Patten}, {Polizotti}, {Smith}, {Taylor}, {Wang}, {Willner}, {Hoffmann},
  {Pipher}, {Forrest}, {McMurty}, {McCreight}, {McKelvey}, {McMurray}, {Koch},
  {Moseley}, {Arendt}, {Mentzell}, {Marx}, {Losch}, {Mayman}, {Eichhorn},
  {Krebs}, {Jhabvala}, {Gezari}, {Fixsen}, {Flores}, {Shakoorzadeh}, {Jungo},
  {Hakun}, {Workman}, {Karpati}, {Kichak}, {Whitley}, {Mann}, {Tollestrup},
  {Eisenhardt}, {Stern}, {Gorjian}, {Bhattacharya}, {Carey}, {Nelson},
  {Glaccum}, {Lacy}, {Lowrance}, {Laine}, {Reach}, {Stauffer}, {Surace},
  {Wilson}, {Wright}, {Hoffman}, {Domingo}, \& {Cohen}}]{2004ApJS..154...10F}
{Fazio}, G.~G., {Hora}, J.~L., {Allen}, L.~E., {et~al.} 2004, \apjs, 154, 10

\bibitem[{{Fesen} {et~al.}(1985){Fesen}, {Blair}, \&
  {Kirshner}}]{1985ApJ...292...29F}
{Fesen}, R.~A., {Blair}, W.~P., \& {Kirshner}, R.~P. 1985, \apj, 292, 29

\bibitem[{{Filipovic} {et~al.}(1998){Filipovic}, {Haynes}, {White}, \&
  {Jones}}]{1998A&AS..130..421F}
{Filipovic}, M.~D., {Haynes}, R.~F., {White}, G.~L., \& {Jones}, P.~A. 1998,
  \aaps, 130, 421

\bibitem[{{Fukui} {et~al.}(2008){Fukui}, {Kawamura}, {Minamidani}, {Mizuno},
  {Kanai}, {Mizuno}, {Onishi}, {Yonekura}, {Mizuno}, {Ogawa}, \&
  {Rubio}}]{2008ApJS..178...56F}
{Fukui}, Y., {Kawamura}, A., {Minamidani}, T., {et~al.} 2008, \apjs, 178, 56

\bibitem[{{Grondin} {et~al.}(2012){Grondin}, {Sasaki}, {Haberl}, {Pietsch},
  {Crawford}, {Filipovi{\'c}}, {Bozzetto}, {Points}, \&
  {Smith}}]{2012A&A...539A..15G}
{Grondin}, M.-H., {Sasaki}, M., {Haberl}, F., {et~al.} 2012, \aap, 539, A15

\bibitem[{{Gruendl} \& {Chu}(2009)}]{2009ApJS..184..172G}
{Gruendl}, R.~A. \& {Chu}, Y.-H. 2009, \apjs, 184, 172

\bibitem[{{Haberl} \& {Pietsch}(1999)}]{1999A&AS..139..277H}
{Haberl}, F. \& {Pietsch}, W. 1999, \aaps, 139, 277

\bibitem[{{Harris} \& {Zaritsky}(2008)}]{2008PASA...25..116H}
{Harris}, J. \& {Zaritsky}, D. 2008, \pasa, 25, 116

\bibitem[{{Henize}(1956)}]{1956ApJS....2..315H}
{Henize}, K.~G. 1956, \apjs, 2, 315

\bibitem[{{Henley} \& {Shelton}(2008)}]{2008ApJ...676..335H}
{Henley}, D.~B. \& {Shelton}, R.~L. 2008, \apj, 676, 335

\bibitem[{{H{\o}g} {et~al.}(2000){H{\o}g}, {Fabricius}, {Makarov}, {Urban},
  {Corbin}, {Wycoff}, {Bastian}, {Schwekendiek}, \&
  {Wicenec}}]{2000A&A...355L..27H}
{H{\o}g}, E., {Fabricius}, C., {Makarov}, V.~V., {et~al.} 2000, \aap, 355, L27

\bibitem[{{Hughes} {et~al.}(2007){Hughes}, {Staveley-Smith}, {Kim}, {Wolleben},
  \& {Filipovi{\'c}}}]{2007MNRAS.382..543H}
{Hughes}, A., {Staveley-Smith}, L., {Kim}, S., {Wolleben}, M., \&
  {Filipovi{\'c}}, M. 2007, \mnras, 382, 543

\bibitem[{{Hughes} {et~al.}(1998){Hughes}, {Hayashi}, \&
  {Koyama}}]{1998ApJ...505..732H}
{Hughes}, J.~P., {Hayashi}, I., \& {Koyama}, K. 1998, \apj, 505, 732

\bibitem[{{Klimek} {et~al.}(2010){Klimek}, {Points}, {Smith}, {Shelton}, \&
  {Williams}}]{2010ApJ...725.2281K}
{Klimek}, M.~D., {Points}, S.~D., {Smith}, R.~C., {Shelton}, R.~L., \&
  {Williams}, R. 2010, \apj, 725, 2281

\bibitem[{{Koo} {et~al.}(2007){Koo}, {Lee}, {Moon}, {Lee}, {Seok}, {Lee},
  {Hong}, {Lee}, {Kaneda}, {Ita}, {Jeong}, {Onaka}, {Sakon}, {Nakagawa}, \&
  {Murakami}}]{2007PASJ...59S.455K}
{Koo}, B.-C., {Lee}, H.-G., {Moon}, D.-S., {et~al.} 2007, \pasj, 59, 455

\bibitem[{{Kuntz} \& {Snowden}(2008)}]{2008A&A...478..575K}
{Kuntz}, K.~D. \& {Snowden}, S.~L. 2008, \aap, 478, 575

\bibitem[{{Kuntz} \& {Snowden}(2010)}]{2010ApJS..188...46K}
{Kuntz}, K.~D. \& {Snowden}, S.~L. 2010, \apjs, 188, 46

\bibitem[{{Lejeune} \& {Schaerer}(2001)}]{2001A&A...366..538L}
{Lejeune}, T. \& {Schaerer}, D. 2001, \aap, 366, 538

\bibitem[{{Lucke} \& {Hodge}(1970)}]{1970AJ.....75..171L}
{Lucke}, P.~B. \& {Hodge}, P.~W. 1970, \aj, 75, 171

\bibitem[{{Mathewson} \& {Clarke}(1973)}]{1973ApJ...180..725M}
{Mathewson}, D.~S. \& {Clarke}, J.~N. 1973, \apj, 180, 725

\bibitem[{{Meixner} {et~al.}(2006){Meixner}, {Gordon}, {Indebetouw}, {Hora},
  {Whitney}, {Blum}, {Reach}, {Bernard}, {Meade}, {Babler}, {Engelbracht},
  {For}, {Misselt}, {Vijh}, {Leitherer}, {Cohen}, {Churchwell}, {Boulanger},
  {Frogel}, {Fukui}, {Gallagher}, {Gorjian}, {Harris}, {Kelly}, {Kawamura},
  {Kim}, {Latter}, {Madden}, {Markwick-Kemper}, {Mizuno}, {Mizuno}, {Mould},
  {Nota}, {Oey}, {Olsen}, {Onishi}, {Paladini}, {Panagia}, {Perez-Gonzalez},
  {Shibai}, {Sato}, {Smith}, {Staveley-Smith}, {Tielens}, {Ueta}, {van Dyk},
  {Volk}, {Werner}, \& {Zaritsky}}]{2006AJ....132.2268M}
{Meixner}, M., {Gordon}, K.~D., {Indebetouw}, R., {et~al.} 2006, \aj, 132, 2268

\bibitem[{{Meynet} {et~al.}(1994){Meynet}, {Maeder}, {Schaller}, {Schaerer}, \&
  {Charbonnel}}]{1994A&AS..103...97M}
{Meynet}, G., {Maeder}, A., {Schaller}, G., {Schaerer}, D., \& {Charbonnel}, C.
  1994, \aaps, 103, 97

\bibitem[{{Micelotta} {et~al.}(2010{\natexlab{a}}){Micelotta}, {Jones}, \&
  {Tielens}}]{2010A&A...510A..37M}
{Micelotta}, E.~R., {Jones}, A.~P., \& {Tielens}, A.~G.~G.~M.
  2010{\natexlab{a}}, \aap, 510, A37

\bibitem[{{Micelotta} {et~al.}(2010{\natexlab{b}}){Micelotta}, {Jones}, \&
  {Tielens}}]{2010A&A...510A..36M}
{Micelotta}, E.~R., {Jones}, A.~P., \& {Tielens}, A.~G.~G.~M.
  2010{\natexlab{b}}, \aap, 510, A36

\bibitem[{{Mills} {et~al.}(1984){Mills}, {Turtle}, {Little}, \&
  {Durdin}}]{1984AuJPh..37..321M}
{Mills}, B.~Y., {Turtle}, A.~J., {Little}, A.~G., \& {Durdin}, J.~M. 1984,
  Australian Journal of Physics, 37, 321

\bibitem[{{Rieke} {et~al.}(2004){Rieke}, {Young}, {Engelbracht}, {Kelly},
  {Low}, {Haller}, {Beeman}, {Gordon}, {Stansberry}, {Misselt}, {Cadien},
  {Morrison}, {Rivlis}, {Latter}, {Noriega-Crespo}, {Padgett}, {Stapelfeldt},
  {Hines}, {Egami}, {Muzerolle}, {Alonso-Herrero}, {Blaylock}, {Dole}, {Hinz},
  {Le Floc'h}, {Papovich}, {P{\'e}rez-Gonz{\'a}lez}, {Smith}, {Su}, {Bennett},
  {Frayer}, {Henderson}, {Lu}, {Masci}, {Pesenson}, {Rebull}, {Rho}, {Keene},
  {Stolovy}, {Wachter}, {Wheaton}, {Werner}, \&
  {Richards}}]{2004ApJS..154...25R}
{Rieke}, G.~H., {Young}, E.~T., {Engelbracht}, C.~W., {et~al.} 2004, \apjs,
  154, 25

\bibitem[{{Russell} \& {Dopita}(1992)}]{1992ApJ...384..508R}
{Russell}, S.~C. \& {Dopita}, M.~A. 1992, \apj, 384, 508

\bibitem[{{Sanduleak}(1970)}]{1970CoTol..89.....S}
{Sanduleak}, N. 1970, Contributions from the Cerro Tololo Inter-American
  Observatory, 89

\bibitem[{{Smith} {et~al.}(2000){Smith}, {Leiton}, \&
  {Pizarro}}]{2000ASPC..221...83S}
{Smith}, C., {Leiton}, R., \& {Pizarro}, S. 2000, in Astronomical Society of
  the Pacific Conference Series, Vol. 221, Stars, Gas and Dust in Galaxies:
  Exploring the Links, ed. {D.~Alloin, K.~Olsen, \& G.~Galaz}, 83

\bibitem[{{Str{\"u}der} {et~al.}(2001){Str{\"u}der}, {Briel}, {Dennerl},
  {Hartmann}, {Kendziorra}, {Meidinger}, {Pfeffermann}, {Reppin}, {Aschenbach},
  {Bornemann}, {Br{\"a}uninger}, {Burkert}, {Elender}, {Freyberg}, {Haberl},
  {Hartner}, {Heuschmann}, {Hippmann}, {Kastelic}, {Kemmer}, {Kettenring},
  {Kink}, {Krause}, {M{\"u}ller}, {Oppitz}, {Pietsch}, {Popp}, {Predehl},
  {Read}, {Stephan}, {St{\"o}tter}, {Tr{\"u}mper}, {Holl}, {Kemmer}, {Soltau},
  {St{\"o}tter}, {Weber}, {Weichert}, {von Zanthier}, {Carathanassis}, {Lutz},
  {Richter}, {Solc}, {B{\"o}ttcher}, {Kuster}, {Staubert}, {Abbey}, {Holland},
  {Turner}, {Balasini}, {Bignami}, {La Palombara}, {Villa}, {Buttler},
  {Gianini}, {Lain{\'e}}, {Lumb}, \& {Dhez}}]{2001A&A...365L..18S}
{Str{\"u}der}, L., {Briel}, U., {Dennerl}, K., {et~al.} 2001, \aap, 365, L18

\bibitem[{{Tappe} {et~al.}(2006){Tappe}, {Rho}, \&
  {Reach}}]{2006ApJ...653..267T}
{Tappe}, A., {Rho}, J., \& {Reach}, W.~T. 2006, \apj, 653, 267

\bibitem[{{Temim} {et~al.}(2006){Temim}, {Gehrz}, {Woodward}, {Roellig},
  {Smith}, {Rudnick}, {Polomski}, {Davidson}, {Yuen}, \&
  {Onaka}}]{2006AJ....132.1610T}
{Temim}, T., {Gehrz}, R.~D., {Woodward}, C.~E., {et~al.} 2006, \aj, 132, 1610

\bibitem[{{Turner} {et~al.}(2001){Turner}, {Abbey}, {Arnaud}, {Balasini},
  {Barbera}, {Belsole}, {Bennie}, {Bernard}, {Bignami}, {Boer}, {Briel},
  {Butler}, {Cara}, {Chabaud}, {Cole}, {Collura}, {Conte}, {Cros}, {Denby},
  {Dhez}, {Di Coco}, {Dowson}, {Ferrando}, {Ghizzardi}, {Gianotti}, {Goodall},
  {Gretton}, {Griffiths}, {Hainaut}, {Hochedez}, {Holland}, {Jourdain},
  {Kendziorra}, {Lagostina}, {Laine}, {La Palombara}, {Lortholary}, {Lumb},
  {Marty}, {Molendi}, {Pigot}, {Poindron}, {Pounds}, {Reeves}, {Reppin},
  {Rothenflug}, {Salvetat}, {Sauvageot}, {Schmitt}, {Sembay}, {Short},
  {Spragg}, {Stephen}, {Str{\"u}der}, {Tiengo}, {Trifoglio}, {Tr{\"u}mper},
  {Vercellone}, {Vigroux}, {Villa}, {Ward}, {Whitehead}, \&
  {Zonca}}]{2001A&A...365L..27T}
{Turner}, M.~J.~L., {Abbey}, A., {Arnaud}, M., {et~al.} 2001, \aap, 365, L27

\bibitem[{{van der Heyden} {et~al.}(2004){van der Heyden}, {Bleeker}, \&
  {Kaastra}}]{2004A&A...421.1031V}
{van der Heyden}, K.~J., {Bleeker}, J.~A.~M., \& {Kaastra}, J.~S. 2004, \aap,
  421, 1031

\bibitem[{{Vanhala} \& {Cameron}(1998)}]{1998ApJ...508..291V}
{Vanhala}, H.~A.~T. \& {Cameron}, A.~G.~W. 1998, \apj, 508, 291

\bibitem[{{Watson} {et~al.}(2009){Watson}, {Schr{\"o}der}, {Fyfe}, {Page},
  {Lamer}, {Mateos}, {Pye}, {Sakano}, {Rosen}, {Ballet}, {Barcons}, {Barret},
  {Boller}, {Brunner}, {Brusa}, {Caccianiga}, {Carrera}, {Ceballos}, {Della
  Ceca}, {Denby}, {Denkinson}, {Dupuy}, {Farrell}, {Fraschetti}, {Freyberg},
  {Guillout}, {Hambaryan}, {Maccacaro}, {Mathiesen}, {McMahon}, {Michel},
  {Motch}, {Osborne}, {Page}, {Pakull}, {Pietsch}, {Saxton}, {Schwope},
  {Severgnini}, {Simpson}, {Sironi}, {Stewart}, {Stewart}, {Stobbart}, {Tedds},
  {Warwick}, {Webb}, {West}, {Worrall}, \& {Yuan}}]{2009A&A...493..339W}
{Watson}, M.~G., {Schr{\"o}der}, A.~C., {Fyfe}, D., {et~al.} 2009, \aap, 493,
  339

\bibitem[{{Williams} {et~al.}(2006){Williams}, {Borkowski}, {Reynolds},
  {Blair}, {Ghavamian}, {Hendrick}, {Long}, {Points}, {Raymond}, {Sankrit},
  {Smith}, \& {Winkler}}]{2006ApJ...652L..33W}
{Williams}, B.~J., {Borkowski}, K.~J., {Reynolds}, S.~P., {et~al.} 2006, \apjl,
  652, L33

\bibitem[{{Williams} {et~al.}(2004){Williams}, {Chu}, {Dickel}, {Gruendl},
  {Shelton}, {Points}, \& {Smith}}]{2004ApJ...613..948W}
{Williams}, R.~M., {Chu}, Y.-H., {Dickel}, J.~R., {et~al.} 2004, \apj, 613, 948

\bibitem[{{Wilms} {et~al.}(2000){Wilms}, {Allen}, \&
  {McCray}}]{2000ApJ...542..914W}
{Wilms}, J., {Allen}, A., \& {McCray}, R. 2000, \apj, 542, 914

\bibitem[{{Yamaguchi} {et~al.}(2010){Yamaguchi}, {Sawada}, \&
  {Bamba}}]{2010ApJ...715..412Y}
{Yamaguchi}, H., {Sawada}, M., \& {Bamba}, A. 2010, \apj, 715, 412

\bibitem[{{Zaritsky} {et~al.}(2004){Zaritsky}, {Harris}, {Thompson}, \&
  {Grebel}}]{2004AJ....128.1606Z}
{Zaritsky}, D., {Harris}, J., {Thompson}, I.~B., \& {Grebel}, E.~K. 2004, \aj,
  128, 1606

\end{thebibliography}
\end{document}